\documentclass[fleqn,usenatbib]{mnras}

\usepackage{newtxtext,newtxmath}
\usepackage{mathptmx}
\usepackage[T1]{fontenc}
\usepackage{ae,aecompl}
\mathcode`\/="202F

\usepackage{graphicx}    % Including figure files
\usepackage{amsmath}    % Advanced maths commands
\usepackage{lscape}
\usepackage{hyperref}

\newcommand{\feh}{$\mbox{\rm [Fe/H]}$}
\newcommand{\alphafe}{${\rm \left[\alpha/Fe\right]}$}
\newcommand{\vhelio}{$V_{\rm HELIO}$}

\newcommand{\jksp}{(J-K$_{\rm s}$) }
\newcommand{\vip}{(V-I)}
\newcommand{\bvp}{(B-V)}
\newcommand{\ks}{K$_{\rm s}$}

\newcommand{\kms}{km s$^{-1}$}
\newcommand{\masyr}{mas yr$^{-1}$}
\newcommand{\pmra}{$\mu_\alpha$ $\cos{\delta}$}
\newcommand{\pmdec}{$\mu_\delta$}

\newcommand{\teff}{T$_{\rm eff}\ $}
\newcommand{\logg}{$\log{ g}\ $}

\newcommand{\mgfe}{$\mbox{\rm [Mg/Fe]}$}
\newcommand{\alfe}{$\mbox{\rm [Al/Fe]}$}
\newcommand{\cfe}{$\mbox{\rm [C/Fe]}$}
\newcommand{\nfe}{$\mbox{\rm [N/Fe]}$}
\newcommand{\ofe}{$\mbox{\rm [O/Fe]}$}
\newcommand{\nafe}{$\mbox{\rm [Na/Fe]}$}
\newcommand{\sife}{$\mbox{\rm [Si/Fe]}$}
\newcommand{\kfe}{$\mbox{\rm [K/Fe]}$}
\newcommand{\cafe}{$\mbox{\rm [Ca/Fe]}$}
\newcommand{\ndfe}{$\mbox{\rm [Nd/Fe]}$}
\newcommand{\cefe}{$\mbox{\rm [Ce/Fe]}$}
\newcommand{\xfe}{$\mbox{\rm [X/Fe]}$}

\title[APOGEE view of the GC NGC~6544]{APOGEE view of the globular cluster NGC~6544}

\author[F. Gran, M. Zoccali, et al.]{
F. Gran$^{1,2,3}$,
M. Zoccali$^{1,2}$,
A. Rojas-Arriagada$^{1,2}$,
I. Saviane$^{3}$,
R. Contreras Ramos$^{1,2}$,
\newauthor{
R. Beaton$^{4}$,
D. Bizyaev$^{5,6}$,
R. E. Cohen$^{7}$, 
J. G. Fern\'andez-Trincado$^{8,9,10}$,}
\newauthor{
D. A. Garc\'ia-Hern\'andez$^{11,12}$,}
D. Geisler$^{13,14,15}$,
R. R. Lane$^{8,10}$,
D. Minniti$^{16,17}$, 
\newauthor{
C. Moni Bidin$^{18}$,
C. Nitschelm$^{19}$,
J. Olivares Carvajal$^{1,2}$,
K. Pan$^{5}$,
F. I. Rojas$^{1,2}$,
S. Villanova$^{13}$}
\\
$\vspace{0.05cm}$\\
$^{1}$Instituto de Astrof\'isica, Av. Vicu\~na Mackenna 4860, Santiago, Chile\\
$^{2}$Instituto Milenio de Astrof\'isica, Santiago, Chile\\
$^{3}$European Southern Observatory, Alonso de C\'ordova 3107, Santiago, Chile\\
$^{4}$Department of Astrophysical Sciences, Princeton University, 4 Ivy Lane, Princeton, NJ 08544\\
$^{5}$Apache Point Observatory and New Mexico State University, PO. Box 59, Sunspot, NM, 88349-0059, USA\\
$^{6}$Sternberg Astronomical Institute, Moscow State University, Moscow, Russia\\
$^{7}$Space Telescope Science Telescope Institute, 3700 San Martin Drive, Baltimore, MD 21218, USA \\
$^{8}$Instituto de Astronom\'ia y Ciencias Planetarias, Universidad de Atacama, Copayapu 485, Copiap\'o, Chile\\
$^{9}$Institut Utinam, CNRS UMR 6213, Universit\'e Bourgogne-Franche-Comt\'e, OSU THETA Franche-Comt\'e, \\Observatoire de Besan\c{c}on, BP 1615, 25010 Besan\c{c}on Cedex, France\\
$^{10}$Centro de Investigaci\'on en Astronom\'ia, Universidad Bernardo O Higgins, Avenida Viel 1497, Santiago, Chile\\
$^{11}$Instituto de Astrof\'isica de Canarias (IAC), E-38205 La Laguna, Tenerife, Spain\\
$^{12}$Universidad de La Laguna (ULL), Departamento de Astrof\'isica, E-38206 La Laguna, Tenerife, Spain\\
$^{13}$Departamento de Astronom\'ia, Casilla 160-C, Universidad de Concepc\'ion, Concepci\'on, Chile\\
$^{14}$Instituto de Investigaci\'on Multidisciplinario en Ciencia y Tecnolog\'ia, Universidad de La Serena, Avenida Ra\'ul Bitr\'an S/N, La Serena, Chile\\
$^{15}$Departamento de Astronom\'ia, Facultad de Ciencias, Universidad de La Serena, Av. Juan Cisternas 1200, La Serena, Chile\\
$^{16}$Depto. de Ciencias F\'isicas, Facultad de Ciencias Exactas, Universidad Andr\'es Bello, Av. Fern\'andez Concha 700, Las Condes, Santiago, Chile\\
$^{17}$Vatican Observatory, V00120 Vatican City State, Italy\\
$^{18}$Instituto de Astronom\'ia, Universidad Cat\'olica del Norte, Avenida Angamos 0610, Antofagasta, Chile\\
$^{19}$Centro de Astronom\'ia (CITEVA), Universidad de Antofagasta, Avenida Angamos 601, Antofagasta 1270300, Chile\\
}

\date{Accepted 2021 April 02. Received 2021 March 02; in original form 2020 August 20}

\pubyear{2020}
\begin{document}
\label{firstpage}
\pagerange{\pageref{firstpage}--\pageref{lastpage}}
\maketitle

\begin{abstract}
%It should be a single paragraph not more than 250 words (200 words for Letters).
The second phase of the APOGEE survey is providing near-infrared, high-resolution, high signal-to-noise spectra of stars in the halo, disk, bar and bulge of the Milky Way. 
The near-infrared spectral window is especially important in the study of the Galactic bulge, where stars are obscured by the dust and gas of the disk in its line-of-sight. 
We present a chemical characterisation of the globular cluster NGC~6544 with high-resolution spectroscopy. 
The characterisation of the cluster chemical fingerprint, given its status of ``interloper'' towards the Galactic bulge 
and clear signatures of tidal disruption in its core is crucial for future chemical tagging efforts.
Cluster members were selected from the DR16 of the APOGEE survey, using chemo-dynamical criteria of individual stars.
A sample of 23 members of the cluster was selected. An analysis considering the intra-cluster abundance variations, known anticorrelations is given.
According to the RGB content of the cluster, the iron content and $\alpha$-enhancement are \feh\ $= -1.44 \pm 0.04$ dex and \alphafe\ $= 0.20 \pm 0.04$ dex, respectively. 
Cluster members show a significant spread in \feh\ and \alfe\ that is larger than expected based on measurement errors. 
An \alfe\ spread, signal of an Mg-Al anticorrelation is observed and used to constraint the cluster mass budget, along with C, N, Mg, Si, K, Ca, and Ce element variations are discussed.
Across all the analysed evolutionary stages (RGB and AGB), about $\sim2/3$ (14 out of 23) show distinct chemical patterns, possibly associated with second-generation stars.
\end{abstract}


%%%%%%%%%%%%%%%%%%%%%%%%%%%%%%%%%%%%%%%%%%%%%%%%%%

%%%%%%%%%%%%%%%%% BODY OF PAPER %%%%%%%%%%%%%%%%%%

\section{Introduction}
As the oldest objects in our Galaxy, Galactic globular clusters (GCs)  play a crucial role in the characterisation of the early phases of its formation.
While most halo clusters have been extensively studied, GCs close to the plane of the Milky Way (MW) have not been equally explored. 
The variable amount of gas and dust and the contamination by disk and bulge stars in the disk and bulge line-of-sights were the main reasons to evade observations in this area. 
Bulge GCs, in particular, that can be associated with this Galactic component based on their positions, kinematics and metallicities \citep[e.g.][]{minniti95, perezvillegas19},
have been avoided until recently, when near-infrared (near-IR) detectors and spectrographs allowed us to reduce the effect of extinction \cite[e.g.][]{valenti07, valenti10, bica16, cohen17, cohen18}.

NGC~6544 \citep[$\ell=5.83$, $b=-2.20$;][]{cohen14} is  a perfect example  of a
poorly studied  GC, despite being at  a distance of  only $\sim 2.5$ kpc  in the
bulge direction.  The first attempt to characterise NGC~6544 come from 160 
stars measured with photographic plates, \cite{alcaino83} reported a very reddened 
cluster, with E($B-V$) = $0.70$ mag, at a distance of $2.8$ kpc. During the last decade, 
\cite{valenti10} analysed the first near-IR CCD photometry of the cluster and used the slope of the red-giant branch (RGB) to derive a metallicity of \feh=$-1.36$ dex. 
Based on low-resolution spectroscopy, \cite{carretta09c} and \cite{saviane12}
found consistent results of the cluster metallicity with \feh\ of $-1.47$ and 
$-1.43$ dex, respectively, both slightly lower than the photometric determination.  
Recently \cite{cohen14} combined near-IR photometry from the VISTA Variable in 
the V\'ia L\'actea survey \citep[VVV;][]{minniti10, saito12} with optical photometry from 
the Hubble Space Telescope, deriving $d=2.46$ kpc, E($B-V$) = $0.79$ mag, and constraining the 
tidal radius with a lower limit of $r_t = 19$ arcmin. They also concluded that the cluster 
is likely to have undergone a tidal stripping process, under the assumption that NGC~6544 
is a halo cluster currently passing close to the Galactic disk, and projected towards the bulge \citep{bica16}.

Recently, \cite{contreras-ramos17}, \cite{perezvillegas19}, and \cite{massari19} confirmed that the cluster origin was not the bulge based on absolute proper motions (PMs), 
from which they derive the orbit of NGC~6544. Interestingly, \cite{contreras-ramos17} found a remarkable elongation of the shape of the cluster in the direction of the Galactic centre, 
unfortunately without any tidal tail signature.
Finally, and as a side product of the numerous efforts to characterise the GC content in the Gaia Data Release 2 (DR2) catalogue 
\citep{brown18,helmi18,lindegreen18}, \cite{vasiliev19}, \cite{baumgardt18}, and \cite{baumgardt19} derived several structural and photometric properties for NGC~6544.
This is how very precise estimates, based solely on cluster members, of the distance, PMs, core, tidal and half-light radius, and mass were derived.

This paper presents the first high-resolution spectral analysis of 23 NGC~6544 
cluster members, to characterise its internal chemical composition and compare it 
with other globular clusters and the bulge field. The structure of this paper is the following: 
section \ref{sec:APOGEEdata} describes the used data, 
section \ref{sec:NGC6544inAPOGEE} describes cluster members selection process, 
section \ref{sec:fundamentalparameters} provides the metallicity determination for the cluster, 
section \ref{sec:abundances} refers to the chemical abundances of the individual stars and its variations within the cluster,
section \ref{sec:gcontext} gives a context of our findings within a sample of GCs, and finally, 
section \ref{sec:summary} presents our concluding remarks.

\section{APOGEE survey data}
\label{sec:APOGEEdata}

The APOGEE  survey \citep{majewski17} is one  of the projects carried out
within  the  SDSS-III  \citep{eisenstein11} and  the  SDSS-IV  \citep{blanton17}
Collaboration.   APOGEE  uses  a multi-fiber  spectrograph  \citep{wilson10, wilson12, wilson19}  to
obtain high-resolution  ($R \sim 22,000$)  high-SNR ($>$100) spectra  in near-IR
wavelengths (H band; $1.51 - 1.69 \mu$m)  in order to observe all the components
of the MW (halo, disk, bar,  and bulge) from the Apache Point Observatory
\citep[APO;][]{gunn06} and Las Campanas Observatory \citep[LCO;][]{duPont}. 
The survey targets mainly giant stars imposing a colour cut in the dereddened near-IR colour-magnitude diagram  \citep{zasowski13, zasowski17}. 
After the observations, all the data collected are processed by the APOGEE pipeline
\citep{nidever15} to extract and co-add multiple visits of the same
target and calculate accurate radial velocities to the level of $0.1$ \kms\ for
an SNR $> 20$ star with at least three observations.  Finally, all the  individual and co-added
spectra are processed by the APOGEE stellar parameter and chemical abundance
pipeline  \citep[ASPCAP;][]{garciaperez16} to derive abundances for up to 26 chemical
elements \citep[][Smith et al. in prep.]{linelist} in the publicly available Data Release 16 \citep[DR16;][]{DR16}. 
The DR16 version of the APOGEE catalogue is presented by \cite{jonsson20}, and it includes
data for more than $\sim437,000$ stars across all the components in the MW.

%_________________________________________
\section{NGC~6544 in the APOGEE DR16}
\label{sec:NGC6544inAPOGEE}

\begin{figure*}
    \centering
    \includegraphics[scale=0.5]{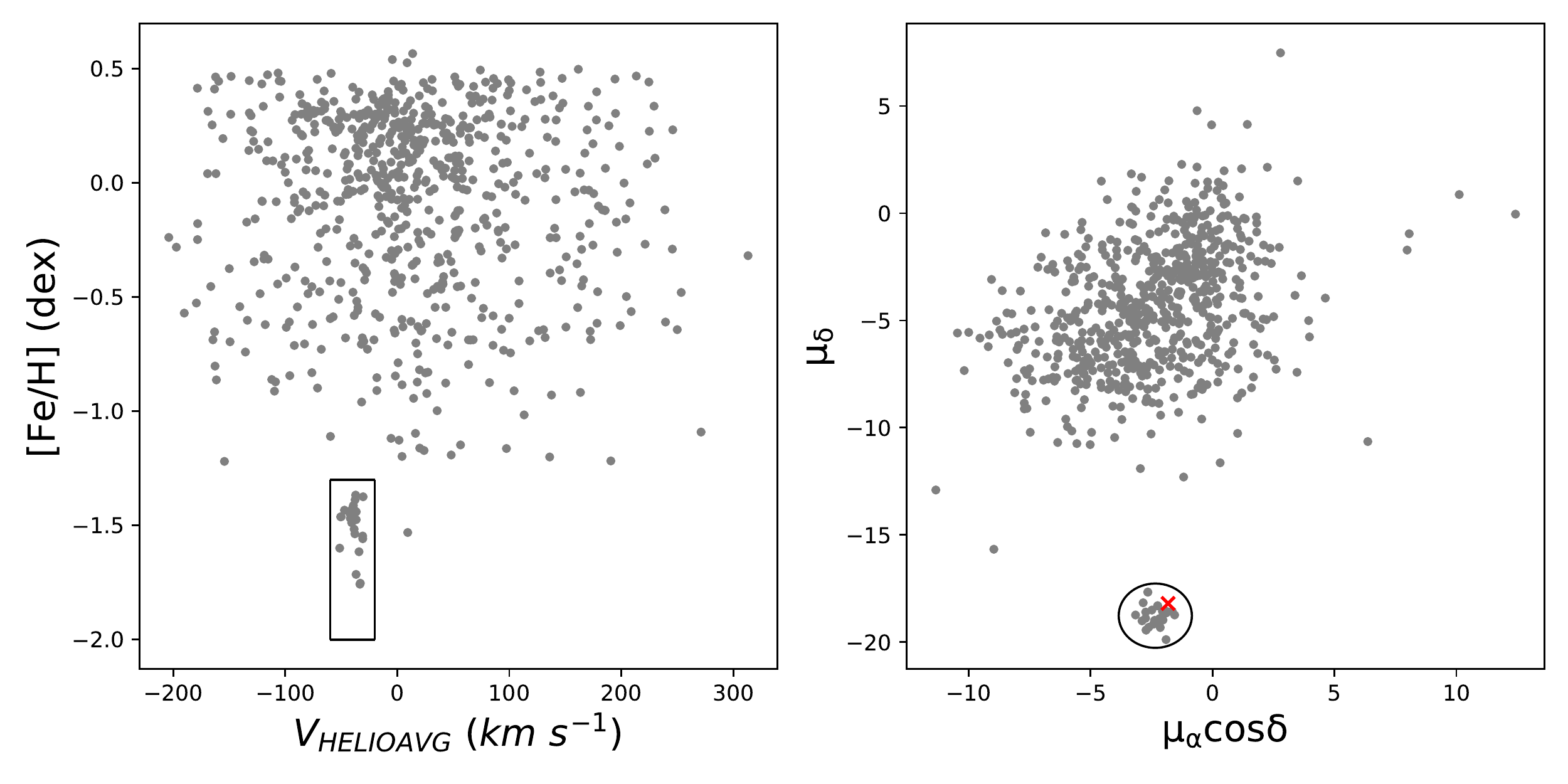}
    \caption{({\bf Left panel}): Heliocentric radial velocity versus \feh\ for all the APOGEE targets within 45 arcmin from 
    the centre of NGC~6544. A group of stars, candidate cluster members, clearly separate from the bulk of field (bulge) stars. See text for the box limits. 
    ({\bf Right panel}): VPD for the same field around the cluster. A small concentration of stars can be seen as a coherent group isolated from the bulge field stars. 
    We choose the radius of the circle as 1.5 \masyr. Note that the RR Lyrae star used to derive the distance to the cluster is marked with a cross.}
    \label{fig:VFe}
\end{figure*}

\begin{figure}
    \centering
    \includegraphics[scale=0.35]{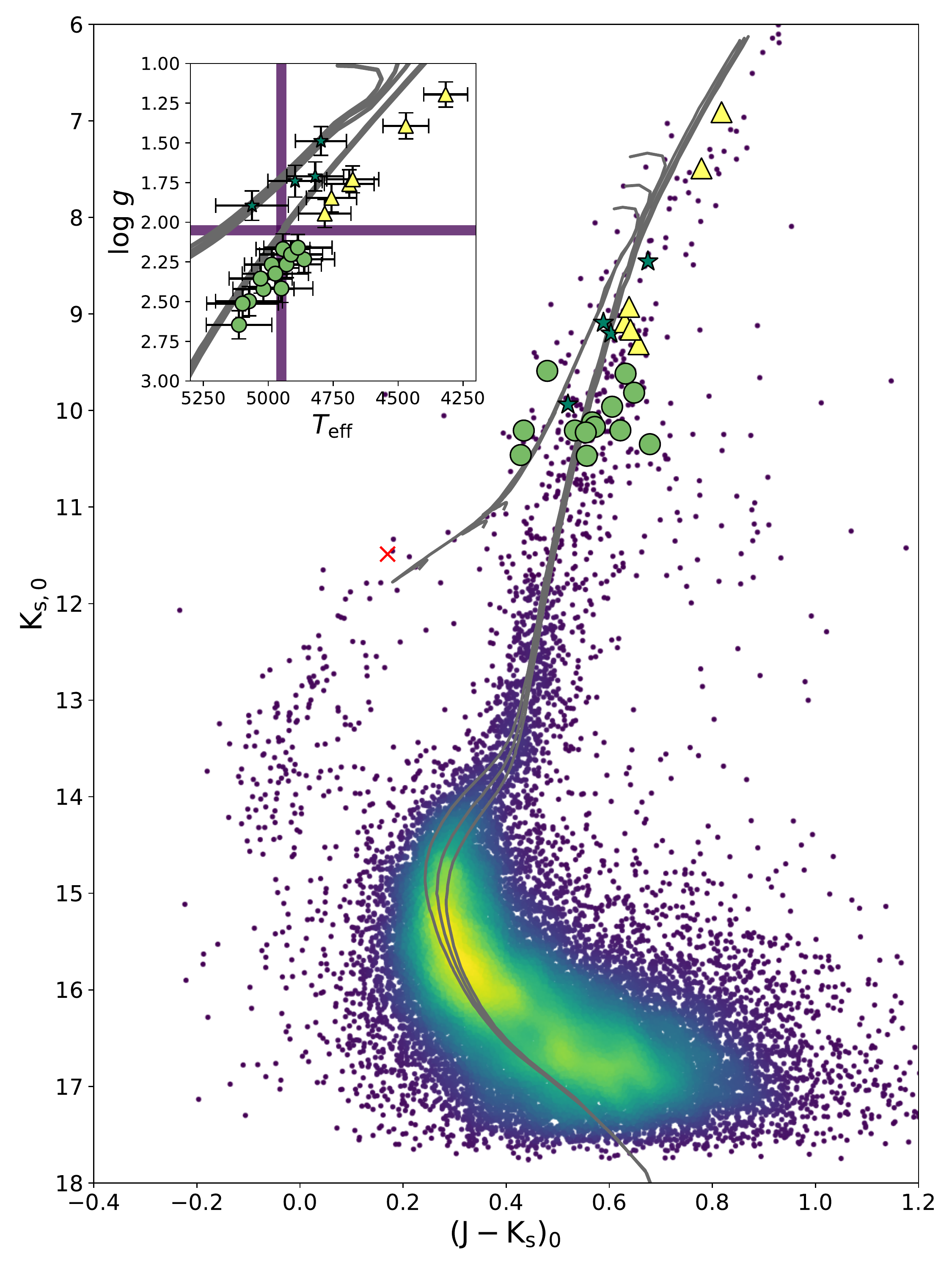}
    \caption{Dereddened \jksp $\times$ \ks color-magnitude diagram of NGC~6544 members from VVV 
    \citep[\ks $\gtrsim 11$ mag,][]{contreras-ramos17} and 2MASS \citep[\ks $\lesssim 11$ mag][]{skrutskie06, valenti07}. 
    As mentioned in the text, different symbols were used to differentiate within stellar stages. 
    % Green circles, yellow triangles, and dark green stars were used to separate lRGB, uRGB, and AGB stars, respectively (see section~\ref{sec:feh} for definitions).
    Circles, triangles, and stars were used to highlight different evolutionary stages based on their positions on the \teff-\logg diagram (see section~\ref{sec:NGC6544inAPOGEE} for definitions).
    Subsequent figures will use the same markers. The RR Lyrae star analysed in section~\ref{sec:distance} is marked with a cross.
    ({\bf Inset}): \teff-\logg diagram of the 23 APOGEE cluster candidates together with the same three PARSEC isochrones. 
    All of them for the same metallicity and $\alpha$-enhancement values of the cluster, for ages of 10, 12 and 14 Gyr. 
    The RGB bump is marked at the intersection of the two shaded areas.}
    \label{fig:cmd}
\end{figure}

We searched for stars that might belong to known globular clusters in the APOGEE public DR16, 
restricted to the bulge area observed by the VVV survey footprint. 
Originally, NGC~6544 was not part of the APOGEE-1 calibration set of clusters (observational campaign form APO), 
and targeting efforts of cluster members started with APOGEE-2 (APO and LCO observations). 
In total, more than 690 stars lie within 45 arcmins of the cluster centre.
The spatial selection cut was performed using the cluster centre derived by \cite{cohen14}, 
as listed in Table~\ref{tab:center}, which centre is shifted by 7 arcsecs with respect to the value provided by the \cite[][2010 edition]{harris96} catalogue. 
Most of the stars within this radius are bulge stars, although 26 stars group around a \vhelio\ 
and \feh\ loci far from the main field distribution, as shown in Figure~\ref{fig:VFe} (left panel).
Note that of the 26 stars selected, there are three with repeated observations from both hemispheres (APO and LCO).

From here on, we will merge the entries selecting the ones with lower overall errors (APO). 
Shown here are the heliocentric velocity and iron abundance derived from the APOGEE reduction pipeline and ASPCAP, respectively. 
The cluster members tend to agglomerate in a region of the \vhelio-\feh\ space marked with a box with 
\feh${}_{\rm median} = -1.46$ dex and \vhelio$_{\rm median} = -38.17$ \kms.
The radial velocity limits of the box have been set as the maximum dispersion found in massive GCs 
\citep{mclaughlin05, watkins15a} of $\lesssim$ 20 \kms, while for the iron content, we select all the stars with
$-2.0 \lesssim \feh\ \left({\rm dex}\right) \lesssim -1.4$ because there are no more stars in this range to be considered cluster members.

As an additional test, we crossmatch the same bulge field stars around NGC~6544 with the Gaia DR2
to obtain PMs with the idea of applying a kinematic cut in our sample. 
For all the cluster candidates we found a match which gives us the PMs
($\mu_{\alpha} \cos \delta$ and $\mu_\delta$ in \masyr), which we show as the vector-point diagram (VPD) in Figure~\ref{fig:VFe} (right panel). 
We select a boundary limit of 1.5 \masyr\ to select cluster members within the VPD.
The median PM of the matched APOGEE stars shows a perfect agreement with the value published by \cite{vasiliev19} and \cite{baumgardt19}. 
We want to perform the cluster selection process as blind as possible, therefore we do not put additional parameter constraints.

In total, the same initial 23 stars were selected as probable NGC~6544 members. The final selection of cluster candidates is 
also showed on their position in the dereddened \jksp {\it vs} \ks color-magnitude diagram (CMD), 
using the 2MASS magnitudes \citep{skrutskie06} transformed to the VISTA filters\footnote{\url{http://casu.ast.cam.ac.uk/surveys-projects/vista/technical/photometric-properties}}. 
Figure~\ref{fig:cmd} shows the CMD of NGC~6544 with the selected APOGEE targets shown with symbols. 
For stars with \ks $>$ 12 mag, VVV dereddened magnitudes come from the PM selected cluster members, according to \cite{contreras-ramos17}. 
However for \ks $<$ 12 mag, stars are saturated in the VVV survey, and therefore we perform a dynamical selection (with 1.5 \masyr\ tolerance) based on the Gaia DR2 PMs to 
select 2MASS members \citep{skrutskie06, valenti07}. 
Three isochrones from the PARSEC library \citep{PARSEC1, PARSEC2, PARSEC3, PARSEC4, PARSEC5, PARSEC6}, were overplotted to the data. 
The predefined parameters were used to retrieve the isochrones in the version 1.2S.
We choose ages of 10, 12, and 14 Gyr, that can be noticed from left to right in the turnoff area. 
As expected for the given ages, all the isochrones follow an almost identical line in the main-sequence, RGB and asymptotic giant-branch (AGB).
The same metallicity and $\alpha$-enhancement derived for the cluster (see section \ref{sec:abundances} for the exact values) were used.

The inset in Figure~\ref{fig:cmd} also shows the \teff-\logg diagram for the observed stars with the isochrones mentioned above.
The isochrones help identify the RGB bump in the inset panel, highlighted with two intersecting shaded areas at \teff$\sim 4950$ K and \logg$\sim 2.0$, 
splitting the RGB sample into two. Note that all the stars in the \teff-\logg diagram appear to be shifted to the respective isochrones towards cooler temperatures or higher gravities. 
The shift is small, and similar to the total error on the parameters, with values of $\Delta$\teff$\sim 100 $ K and $\Delta$\logg$\sim 0.2$ dex 
in the uncorrelated case and half of them if we consider a simultaneous shift on both measurements. 
Despite this fact and as expected for a single GC, stars follow the evolutionary sequences, confirming that the stellar surface parameters are rather well determined. 
On the contrary, the targets are more spread out in the observed colour-magnitude diagram because of the large differential reddening along the line-of-sight: 
up to $\Delta$E\jksp $\sim 0.4$ mag within arcmin scales around the cluster centre, as already noted by \citet{contreras-ramos17}.
It is for this reason that we only use the position of the stars in the \teff-\logg diagram to classify the cluster members into three groups:
the main two in the RGB evolutionary phase, below and above the bump and the third one, associated to the AGB phase.
Take in consideration that the RGB-bump might differ whether we select a stellar model or perform an empirical estimate, as shown in \cite{cohen17} for NGC~6544.
The shift in magnitudes can be as large as $\Delta {\rm K_s}\sim 0.2-0.3$ mags for this cluster, depending on the derived distance, $\alpha$-enhancement, and to a lesser extent by age \citep{salaris07}.

As Figure~\ref{fig:cmd} suggests, we split the cluster sample into three groups: 
stars below and above the RGB bump (lower RGB or lRGB as circles, and upper RGB or uRGB as triangles), and the AGB group (as stars).
Each group consists of 13, 6 and 4 stars for the lRGB, uRGB, and AGB, respectively, with a total of 23 observed cluster stars.
These groups were also marked with different symbols according to APOGEE derived \teff\ and \logg\ values.

In section \ref{sec:fundamentalparameters} and \ref{sec:abundances}, abundance differences will be reviewed, but as expected from an evolutionary point of view, stars below the RGB bump will be Carbon-enhanced and Nitrogen-depleted, and the contrary for the stars above the bump.

\begin{table*}
    \centering
    \caption{NGC~6544 coordinates (J2000), tidal radius and distance from the Sun \citep{cohen14}, metallicity \citep[][2010 edition]{harris96} and PMs \citep{baumgardt19}.}
    \label{tab:center}
  \begin{tabular}{c|c|c|c|c|c|c|c|c}
      RA      &      Dec    & $\ell$ &   $b$   & $r_{\rm t}$ & $d_\odot$ & \feh\ &  $\mu_{\alpha} \cos \delta$ & $\mu_\delta$\\
  (hh:mm:ss)  &  (dd:mm:ss) &  (deg) &  (deg)  &   (arcmin)  &   (kpc)   & (dex)  & (\masyr) & (\masyr) \\
  \hline
  18:07:20.12 & -24:59:53.6 & 5.8365 & -2.2024 &    $\simeq 19$    &   2.46 $\pm$ 0.09& -1.4 & -2.34 $\pm$ 0.04 & -18.66 $\pm$ 0.04 \\
  \hline
  \end{tabular}
\end{table*}

\section{NGC~6544 fundamental parameters}
\label{sec:fundamentalparameters}

\subsection{Metallicity and $\alpha$-enhancement values}
\label{sec:fehalpha}

From the photometric point of view, some of the most uncertain parameters to determine from a CMD are the metallicity, $\alpha$-enhancement, and age of a stellar population.
Even a slight change in the first two affects the main-sequence turnoff point location, and therefore the estimated age of a stellar population \citep[e.g.][]{oliveira20,souza20}.
Here the cluster metallicity is known in great detail even across different stages of stellar evolution, and the combination of 
PMs and radial velocities ensures that all the analysed stars move coherently in space and form part of the cluster.

One of the first striking results that the spectra present is the systematic difference between the DR16 derived metallicity of the RGB (both groups) and that of the AGB.
The marked difference between populations of a globular cluster was not present in previous studies, 
while using APOGEE spectra but applying different reduction methods \citep{garciahernandez15, masseron19, meszaros20}.  
In this cluster we are finding that the median metallicity with its standard error for the entire RGB is \feh=$ -1.44 \pm 0.04$ dex, 
that is composed of a lower and upper part with $-1.44 \pm 0.05$ dex and $-1.46 \pm 0.04$ dex, respectively.
RGB metallicities agrees within its errors, nevertheless are systematically higher than those of AGB stars of \feh=$ -1.66 \pm 0.10$ dex. 
Across the paper we use the standard deviation of the median as a statistically significant measurement on the errors.

The discrepant values of iron content for stars in different evolutionary stages (RGB and AGB in this case) 
is not a new finding. Indeed, \citet{ivans01} found that AGB stars in M5 had [FeI/H] lower than RGB stars, by 0.15 dex.
They concluded that this finding was in agreement with the prediction of \citet{thevenin99}, i.e., that in the atmosphere 
of metal poor, late type stars, most FeI lines are formed out of local thermal equilibrium (LTE), and therefore by assuming LTE, derived FeI abundances are lower than real.
This is due to the fact that in thin, transparent atmospheres, some of the Fe is ionised by UV radiation coming from the stellar interior. 
The effect is larger for less dense atmospheres at a given \teff, and also larger for higher \teff at a given luminosity, therefore it is expected to be larger in AGB stars than in RGB stars.
According to \citet{thevenin99}, FeII lines are not affected by this NLTE effect.
More recently and using optical spectra, \citet{lapenna14, lapenna15, mucciarelli15a, mucciarelli15b} all confirmed that the putative spread in the 
iron abundance of stars in 47Tuc, M62, M22, NGC3201, reported in previous studies, was due to the same effect.
Namely, FeI abundances in AGB stars were lower than those of RGB stars. 
On the contrary, iron abundances derived from FeII lines showed a very good agreement in both evolutionary phases. 
However, converging metallicities between the RGB and AGB stages were found by \cite{garciahernandez15,masseron19,meszaros20} when characterising APOGEE near-IR spectra that uses only FeI lines. 
With these conflicting evidence in context, it is clear that we have no clear explanation of the phenomena that NGC~6544 AGB stars exhibit, giving that they are dynamically confirmed cluster members.
For that reason, we have isolated the behaviour of the different evolutionary stages to compute the median \feh.

A max-to-mean difference of 0.23 dex is observed in \feh\ (or 0.17 dex within each RGB subdivision), a value that exceeds the 
observational errors in NGC~6544 (typically $\pm$0.02 dex), and in other GCs \citep{carretta09c}.
It will be evident from Section~\ref{sec:abundances} that the star-to-star variations exceed the 
observational error, even among the RGB and AGB stars themselves.
We cannot exclude that extra-enrichment processes happened in NGC~6544 \citep[as explained in][]{renzini08}, or even that 
this trend is the result of the interaction with the MW \citep{contreras-ramos18a,kundu19a,kundu19b}. 
Given the absolute $V$ magnitude ($M_{V_t}$) of -6.94 mags as a proxy of the total mass, no other cluster with similar $M_{V_t}$
shows an \feh\ dispersion greater than $\sim 0.07$ dex \citep{harris96,bailin19}.
Up to now, we do not have a clear explanation of the origin of the \feh\ spread of NGC~6544, considering the limited number of stars observed.

Finally, the ASPCAP pipeline returns an estimation for the \alphafe\ content of each analysed star.
Following the description in \cite{garciaperez16}, this contribution is measured using O, Mg, Si, S, Ca, and Ti from the spectra.
The median \alphafe\ value obtained from cluster combined RGB stars is $0.20 \pm 0.04$ dex, matching canonical values for GCs. 
If we separate stars in the lower and upper part of the RGB, we find medians of $0.22 \pm 0.04$ dex and $0.19 \pm 0.03$ dex,
respectively. The same deviation from the RGB values is found in the AGB stars, with a median of $0.25 \pm 0.08$ dex.
We present median observed quantities, instead of mean, because of a significant spread in the \feh\ and \alphafe\ of cluster
stars.

\subsection{Distance to NGC~6544}
\label{sec:distance}

Distance is a crucial parameter that is often derived from CMD fitting. Although it degenerates with reddening and metallicity.
A CMD-independent measurement of the distance can be sought to confirm that the isochrone parameters were accurate enough.
Figure~\ref{fig:cmd} uses a reddening value of E\jksp = 0.36 mags \citep{contreras-ramos17} in the VISTA filters or $0.40$ mags 
in the 2MASS system and a distance modulus of $dm = 11.92$ found in this work, 
which is consistent with the distances derived by \cite{cohen14} and \cite{contreras-ramos17} of $dm = 11.96$ mag.

Although only one RR Lyrae star has been detected to move along with the cluster (see Figure~\ref{fig:VFe} right panel), it can be used to verify the assumed distance modulus. 
The RR Lyrae is listed in the online catalogue of \cite{clement}, it was also detected by the Optical Gravitational 
Lensing Experiment (OGLE) survey \cite[OGLE-BLG-RRLYR-13867\footnote{Available at the OGLE website: \url{http://ogledb.astrouw.edu.pl/~ogle/OCVS/o.php?OGLE-BLG-RRLYR-13867}}, ][]{oglerrl,ogleiv}, 
and by the VVV survey (Contreras Ramos private communication). 
Its main properties are presented in Table~\ref{tab:rrl}, yielding a distance modulus of $dm = 11.92 \pm 0.18$ mag 
($2.43 \pm 0.20$ kpc) using the \cite{muraveva15} period-luminosity relation for the K$_{\rm s}$ band, an intrinsic \jksp RR Lyrae colour of 
$0.17 \pm 0.03$ \citep{navarrete15, contreras-ramos18b} and the \cite{cardelli89} extinction law, which is comparable to other distance measurements.
We also report the near-IR reddening and extinction values given by the RR Lyrae star in Table~\ref{tab:rrl}. 
Consider that all the near-IR measurements presented here are in the 2MASS filter system.
Note that E\jksp and E\vip derived are consistent within errors with \cite{contreras-ramos17} and \cite{cohen14}, respectively.
To calculate E\vip, we use the same derivation described by \cite{pietrukowicz15}, considering the actual metallicity of the cluster derived above and the \cite{catelan04} period-luminosity relations.
There is also a discrepancy in the E\jksp value derived by \cite{cohen14} which is $\sim$ 0.1 mags shifted towards 
higher reddening. This shift will be addressed in the next subsection as a VVV survey calibration issue discovered in \cite{hajdu20}.

Finally, \cite{vasiliev21} recently derived a parallax-driven value of $d = 2.54 \pm 0.07$ kpc ($\overline\varpi =  0.394 \pm 0.011$ mas) to NGC~6544.
This is an independent distance determination, which is also in agreement with all of the previous measurements showed in this section. 

\begin{table}
    \centering
    \caption{RR Lyrae star used to verify the NGC~6544 distance modulus. Magnitudes in each filter corresponds to the mean quantity.
    J- and H-band values were derived from only a few epochs, as described in \citet{saito12}.}
    \label{tab:rrl}
    \scriptsize
    \begin{tabular}{c|c}
    \hline
    \multicolumn{2}{|c|}{Stellar parameters from Gaia DR2} \\
    \hline
        Source ID & 4065784375110674176 \\
        RA (J2000) & 271.868215 \\
        Dec (J2000) & -25.031377 \\
        $\ell$ (deg) & 5.8226 \\
        $b$ (deg) & 2.2458 \\
        G (mag) & 14.5874 $\pm$ 0.0311 \\
        BP (mag) & 15.2310 $\pm$ 0.1308 \\
        RP (mag) & 13.6256 $\pm$ 0.0770 \\
        \pmra\ (\masyr) & -1.821$\pm$0.074 \\
        \pmdec (\masyr) & -18.195$\pm$0.061 \\
    \hline
    \multicolumn{2}{|c|}{Stellar parameters from OGLE survey} \\
    \hline
        OGLE ID & OGLE-BLG-RRLYR-13867 \\
        V (mag) & 13.584 \\
        I (mag) & 15.048 \\
        P (days) & 0.5723310 $\pm$ 0.0000001 \\
        $I_{\rm amplitude}$ (mag) & 0.777 \\
    \hline
    \multicolumn{2}{|c|}{Stellar parameters from VVV survey} \\
    \hline
        VVV ID & VVV180728.37-250153.1 \\
        J (mag) & 12.164 $\pm$ 0.02 \\
        H (mag) & 11.932 $\pm$ 0.03 \\
        K$_{\rm s}$ (mag) & 11.694$\pm$0.01 \\
    \hline
    \multicolumn{2}{|c|}{Derived stellar parameters} \\
    \hline
        E\vip & 0.98 $\pm$ 0.07 mag  \\
        E\jksp & 0.31 $\pm$ 0.03 mag  \\
        A$_{\rm K_s}$ & 0.25 $\pm$ 0.02 mag   \\
        $dm$ (mag) & 11.92 $\pm$ 0.18 \\
        Distance (kpc) & 2.43 $\pm$ 0.20 \\
    \hline
  \end{tabular}
\end{table}

\subsection{Reddening Law towards NGC~6544}
\label{sec:reddening}

Given the multi-survey synergies that we have presented until this point, the final quantity that we want to constraint is the reddening law towards NGC~6544.
Using the spectroscopic data that we are analysing, it is possible to derive isochrone-based or synthetic colours for all the stars in the cluster.
We will consider, however, only the RGB stars, given the uncertain nature of the AGB stars as discussed in Section~\ref{sec:fehalpha}.
We crossmatch the star positions to a list of surveys that have observed the area with 1 arcsec tolerance, 
founding that there are observations from the ultraviolet (UV) to the mid-IR for part of the stars in our sample.
Photometric information from SWIFT \citep{swift}, PanStarrs DR2 \citep{panstarrs1,panstarrs2}, Gaia DR2 \citep{brown18,lindegreen18}, 
2MASS \citep{skrutskie06}, Spitzer/GLIMPSE \citep{GLIMPSE1,GLIMPSE2} and AllWISE \citep{WISE,NEOWISE} was collected.
In total, 20 filters (${\rm UVOT/UVW2}$; ${\rm g,r,i,z,y}$; ${\rm G, BP, RP}$; ${\rm J,H,K_s}$; ${\rm \left[3.6\right], \left[4.5\right], \left[5.8\right], \left[8.0\right]}$; 
and ${\rm W1,W2,W3,W4}$) contain fluxes for our stars.
Given the crowded nature of a GC, even with the radius tolerance of 1 arcsec, surveys with large pixel scale will often present incorrect or blended magnitudes for the analysed stars.
Note also, that we retrieve Swift matches for only five stars in the NGC~6544 sample, 
in contrast to all the other surveys in which we obtain a high counterpart recovery percentage ($\sim95$\% or 18 out of 19 possible matches in the worst case).
Considering the relevance of the ${\rm B}$ and ${\rm V}$ filters in the extinction characterisation, 
to obtain these magnitudes we apply a colour transformation present in \cite{tonry12} to the PanStarrs ${\rm g}$ and ${\rm r}$ bandpasses.

To derive synthetic magnitudes we interpolate the \teff\ and \logg from the APOGEE DR16 into a PARSEC isochrone 
of 12 Gyr and the cluster metallicity to obtain the absolute magnitudes in different filters.
With the absolute theoretical magnitudes, we are able to calculate the colour excess term just subtracting to the observed colour term.
This step is necessary to compare with other results, like the one that we derive from the RR Lyrae colours, or the canonical E\bvp\ value.
Indeed, it is important to mention that there should be statistical and systematic errors not considered here, like the isochrone election or additional differential reddening effects.

Our results for the E\bvp\ $=0.77 \pm 0.19$ mag coincides with the one derived by \cite{cohen14} of $0.79 \pm 0.01$ mag.
However, the E($g-r$) $= 1.19 \pm 0.16$ value from our method is greater than previously published values. 
Extinction maps released\footnote{ Available at the website: \url{http://argonaut.skymaps.info/}} by \cite{green14,green19} account 
for a value of E($g-r$) $= 0.58 \pm 0.03$ mag at the cluster distance.

In parallel, \cite{cohen14}, \cite{contreras-ramos17}, and \cite{surot19,surot20} give values of the E\jksp of $0.43 \pm 0.03$ mag, $0.40$ mag, and $0.47 \pm 0.09$ mag respectively, for which we agree with our method.
Is important to note that to produce the best fit for the Figure~\ref{fig:cmd} CMD we use the E\jksp value present in the \cite{contreras-ramos17} article ($0.36$ mags).
This value is provided in the VISTA system of magnitudes, and it is affected by the VISTA-2MASS colour transformations, and the correction to the zero-point calibration published by \cite{hajdu20}.
After transformation to the 2MASS framework, the \cite{contreras-ramos17} value of E\jksp $= 0.40$ mag, is consistent with our results.
The small discrepancy that is still remains of the order of $\sim0.04$ with our E\jksp derived value can be explained by the strong differential reddening of the cluster,
with have amplitudes of up to $\Delta$E\jksp $= 0.33$ mags depending on the spatial location across the cluster area (Contreras Ramos private communication).
Another possible explanation is possibly the recalibration process proposed by \cite{hajdu20}, however, this effect should be an order of magnitude less prominent than the last one.
With this analysis, we have settled the near-IR reddening discrepancy, preferring the value given by \cite{cohen14} and this work of E\jksp $= 0.44 \pm 0.01$.

\begin{figure}
    \centering
    \includegraphics[scale=0.45]{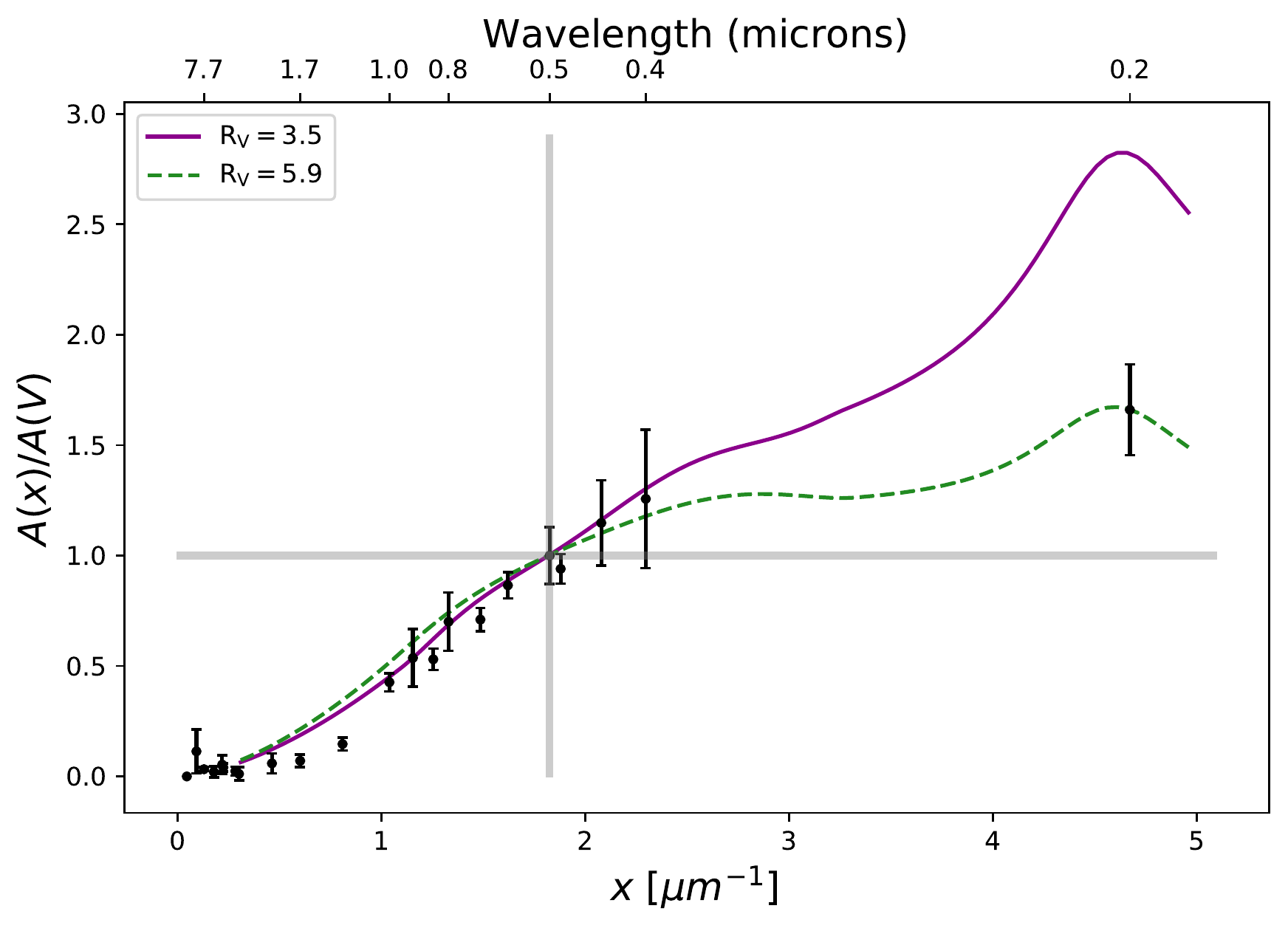}
    \caption{Wavelength (as $x\equiv \lambda^{-1}$[nm$^{-1}$] and $\lambda$ [nm]) against the normalised extinction values with respect to the ${\rm V}$-band ${\rm A}$($x$)$/{\rm A}$(${\rm V}$).
    The two extinction curves shown represents the total-to-selective ratio of $\rm{R_V} = 3.5$ (continuous), and $\rm{R_V} = 5.8$ (dashed) between $0.3$ and $5.0$ nm.
    In total, 20 filters from the UV to the mid-IR with its derives extinction values are shown. 
    The ${\rm V}$-band is highlighted with a shaded area because, by definition, all the extinction curves pass through this point.
    }
        \label{fig:extinction}
\end{figure}

Finally, we take the difference within the observed star magnitudes and the theoretical absolute magnitude, shifted to the cluster distance ($dm = 11.92$ from Table~\ref{tab:rrl}).
The difference corresponds to the total extinction ${\rm A_{\lambda}}$ present in the line of sight per filter, which is presented in Table~\ref{tab:extinctions}.
Giving the positive physical nature of the extinction values, we have not considered some slightly negative results as a product of statistical fluctuations in the data, 
which were probably caused, as we commented earlier, by the differential reddening not taken into account in this step.
As expected, a monotonic decrease trend is observed as the wavelength increases, only interrupted by the reddest filters ${\rm W2}$, $\left[8.0\right]$ and ${\rm W3}$ 
for which we assume that several stars contribute to the matched magnitude in its corresponding catalogues. 
Note also that ${\rm W4}$ shows a very high dispersion compared to the others near- and mid-IR filters that follow the overall trend.
Figure~\ref{fig:extinction} shows the inverse wavelength and normalised extinction values with respect to the ${\rm V}$-band in the same format as in \cite{cardelli89}.
We use the {\tt dust\_extinction} python package to reproduce the MW ${\rm R(V)}$ extinction curve at different ${\rm R_V = A_V/E\left(B-V\right)}$ values.
As it is clear from the Figure, total-to-selective extinction ratio of ${\rm R_V = 3.5 \pm 0.19}$ deduced from the data is enough to 
reproduce the overall optical trend ($0.4 \lesssim \lambda$ (${\rm nm}$) $ \lesssim 1.0$). 
However, this value present inconsistencies with the UV and mid-IR data. 
We attribute this discrepancy to the source confusion caused by the combined effect of the crowding level typical of GCs and the lower spatial resolution (i.e., larger pixel size) of these photometric surveys.
The Figure also shows an extreme case with ${\rm R_V = 5.9}$ that fits solely the UV observation, not reproducing any other available filters to illustrate the the erroneous status of the measurement.
Given the similarities in the optical and IR parts of the spectral energy distribution within the ${\rm R_V = 3.1}$ and ${\rm R_V = 3.5}$ curves of less than $\sim5$\%,
we are not able to differentiate the preferred model, both being consistent with the current data.

As a final caveat on this procedure, all of the extinction derivation relies on the fact that isochrones are model independent, which is not completely accurate in this case.
The PARSEC set of isochrones uses the \cite{cardelli89} recipe to apply the interstellar extinction to their data, adopting a canonical value of ${\rm R_V = 3.1}$.
Nevertheless, given the short distance of the cluster this effect should not represent a dramatic change in our results.

\begin{table}
  \centering
  \caption{ Median extinction coefficients per bandpass derived for the NGC~6544 stars.}
  \label{tab:extinctions}
  \scriptsize
  \begin{tabular}{c|c|c|c|c}
\hline
Filter &        x             & Wavelength &  $A_{\lambda}$ & $A_{\lambda}/A_{\rm V}$ \\
       & ($\mu {\rm m}^{-1}$) &  ($\mu$m)  &     (mag)      &                         \\
\hline
${\rm UVW2}$              & 4.67 &  0.21 & 4.51 $\pm$ 0.90 & 1.69 \\
${\rm B}$                 & 2.30 &  0.44 & 3.42 $\pm$ 0.28 & 1.28 \\
${\rm g}$                 & 2.08 &  0.48 & 3.12 $\pm$ 0.19 & 1.17 \\
${\rm BP}$                & 1.88 &  0.53 & 2.55 $\pm$ 0.10 & 0.96 \\
${\rm V}$                 & 1.83 &  0.55 & 2.72 $\pm$ 0.14 & 1.02 \\
${\rm r}$                 & 1.62 &  0.62 & 2.35 $\pm$ 0.10 & 0.88 \\
${\rm G}$                 & 1.49 &  0.67 & 1.93 $\pm$ 0.08 & 0.72 \\
${\rm i}$                 & 1.33 &  0.75 & 1.90 $\pm$ 0.13 & 0.71 \\
${\rm RP}$                & 1.25 &  0.80 & 1.44 $\pm$ 0.06 & 0.54 \\
${\rm z}$                 & 1.15 &  0.87 & 1.46 $\pm$ 0.12 & 0.55 \\
${\rm y}$                 & 1.04 &  0.96 & 1.16 $\pm$ 0.05 & 0.43 \\
${\rm J}$                 & 0.81 &  1.24 & 0.40 $\pm$ 0.03 & 0.15 \\
${\rm H}$                 & 0.60 &  1.66 & 0.19 $\pm$ 0.03 & 0.07 \\
${\rm K_s}$               & 0.46 &  2.16 & 0.16 $\pm$ 0.03 & 0.06 \\
${\rm W1}$                & 0.30 &  3.32 & 0.03 $\pm$ 0.03 & 0.01 \\
${\rm \left[3.6\right]}$  & 0.28 &  3.52 & 0.06 $\pm$ 0.02 & 0.02 \\
${\rm \left[4.5\right]}$  & 0.22 &  4.45 & 0.11 $\pm$ 0.03 & 0.04 \\
${\rm W2}$                & 0.22 &  4.56 & 0.15 $\pm$ 0.02 & 0.05 \\
${\rm \left[5.8\right]}$  & 0.18 &  5.61 & 0.06 $\pm$ 0.03 & 0.02 \\
${\rm \left[8.0\right]}$  & 0.13 &  7.70 & 0.08 $\pm$ 0.18 & 0.03 \\
${\rm W3}$                & 0.09 & 10.79 & 0.31 $\pm$ 0.04 & 0.12 \\
${\rm W4}$                & 0.05 & 21.91 & 0.00 $\pm$ 0.09 & 0.00 \\
\hline
 \end{tabular}
\end{table}

%_________________________________________
\section{Fingerprint of an elusive cluster}
\label{sec:abundances}

As a result of the earliest star formation in galaxies GCs are one of the main tools that we have to unveil the early stages and evolution of the Milky Way. 
Long considered as simple stellar populations, GCs have revealed a more complex composition of their stars, 
which appears to belong to different generations, each one with star-to-star abundance variations, visible at photometric and spectroscopic scale \cite[][and references therein]{gratton12,bastian18}.

In this sense, these multiple populations (MPs) within a GC refer basically to a first 
(or primordial generation, FG) and at last a second generation (SG) of stars.
The FG of stars pollutes the SG with processed material, and in this way different sub-populations arise in the cluster.
The composition of the SG stars varies almost at an individual cluster level, manifested through several sequences in CMDs, 
as well as light-element variations within the clusters \citep[C, N, O, Na, Mg, Al, Si, among other elements, see][for a recent review]{bastian18}. 

We want to extend this background to NGC~6544 if possible, trying to search for MPs within the member stars through light-element abundances.
In order to interpret NGC~6544 chemical properties, we compare its abundance patterns with other GCs 
observed by APOGEE, and with nearby bulge field stars. All the abundance comparisons were performed with 
the APOGEE DR16 values. Crossmatched stars from other GCs were located in the DR16 using the {\tt APOGEEID} available for all the recent APOGEE publications.

\subsection{Abundance analysis and element variations}

The present analysis adds a substantial contribution to the chemical characterisation of NGC~6544. 
Our sample of cluster members ascends to 23 and allows us to observe intracluster abundance variations comparatively.
% In comparison, \cite{nataf19} and \cite{meszaros20}, only found 7 and 2 stars in the APOGEE DR14 and 16, respectively, 
% because of they rely only on the APOGEE targeting flags and consider the less extincted clusters within the APOGEE sample.
In previous publications, \cite{nataf19} and \cite{meszaros20}, only found 2 and 7 stars in the APOGEE DR14 and 16, respectively.
This difference is originated because of the growing number of observations produced by the APOGEE survey over time, 
and the fact that we not use the GC flag that is present in all the APOGEE DRs ({\tt TARGFLAGS$=$\{APOGEE\_SCI\_CLUSTER, APOGEE2\_SCI\_CLUSTER\}}).
Recently, \cite{horta20} analysed most of the GCs present in the APOGEE catalogue including NGC~6544, in a broad approach 
to compare the accreted and {\it in situ} classification given by \cite{massari19}.
They calculated mean metallicities, radial velocities and \sife\ abundances to derive general GCs properties of the MW, but without detailing any other cluster features. 
The paper indicates that 21 NGC~6544 stars were recovered from the cluster reaching identical results than the ones presented here.
We estimate that the different number of stars analysed originated on the automatic elimination of stars, which removes two of the most metal-poor AGB stars \citep[see Figure 2 in the Appendix of][]{horta20}

Table~\ref{tab:all} contains the abundance values, derived by the ASPCAP pipeline, for a subset of all the elements 
reported in the DR16. Specifically, we select the following elements: \cfe, \nfe, \ofe, \nafe, \mgfe, \alfe, \sife, 
\kfe, and \cafe, together with \alphafe, and \feh, in order to analyse known (anti) correlations for other GCs and to insert
NGC~6544 within the scenario proposed by \citet{jonsson18,zasowski19} and \citet{masseron19} for the bulge and GC samples, 
respectively.

\begin{figure*}
    \centering
    \includegraphics[scale=0.4]{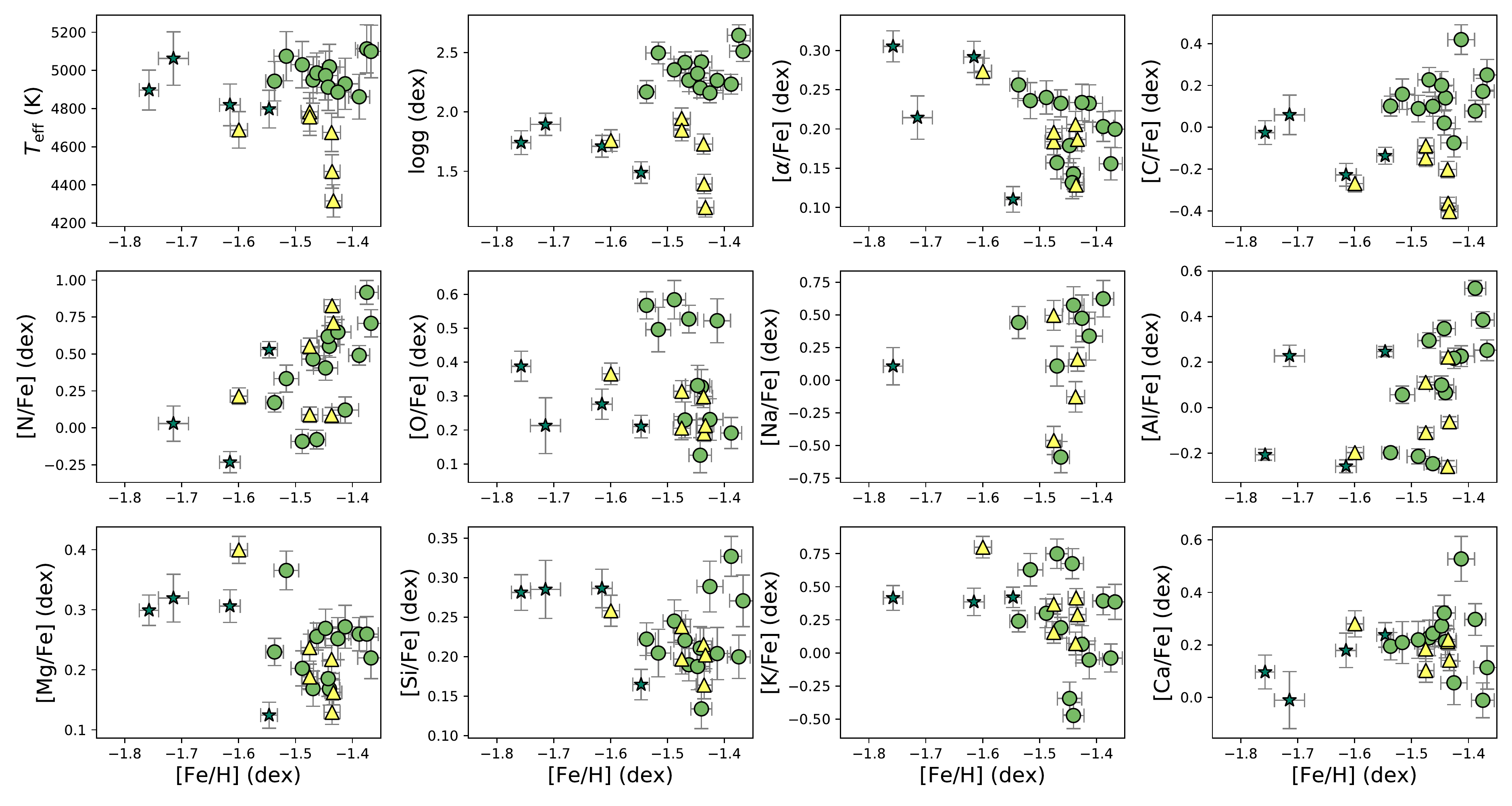}
    \caption{\teff, \logg, and \xfe\ for surface parameters and abundances in Table~\ref{tab:all}. 
    It can be noticed from the Figure, for a given metallicity there is a large spread for certain elements (\nfe, \cfe, \nafe, and \alfe).
    The AGB group (green stars) and one uRGB star (yellow triangle) is completely offset from the global \feh\ trend. 
    This variety of \feh\ values was the main reason which led us to consider and publish the median over the mean values for the elemental abundances.}
    \label{fig:all-vs-fe}
\end{figure*}

Figure~\ref{fig:all-vs-fe} shows the abundance of each element as a function of iron, for all the cluster members. 
Nevertheless, we will base the discussion on RGB stars only, because of the metallicity offset of AGB stars mentioned 
in section~\ref{sec:fundamentalparameters}. Figure~\ref{fig:all-vs-fe} shows a slight trend in sections showing the evolution of \teff, \logg, 
and \alphafe with metallicity for lRGB stars, but the trend vanishes as stars cross the RGB bump.
A clear dichotomy for stars in the lRGB and uRGB 
is present in the \cfe\  abundances as expected from stellar evolution with C-enhanced stars below the RGB bump and 
C-poor stars above the same point \citep{iben68,cassisi12,lardo12}. 

Even if there are only a few Na lines in the APOGEE H-band spectra, only half of the cluster sample has measured 
abundances from the ASPCAP pipeline. They show a wide spread of Na abundance that resemble the trend presented in \cite{masseron19} and \cite{carretta09b}.
A detailed discussion of known (anti)-correlations will be given in the next subsections.

\begin{figure*}
    \centering
    \includegraphics[scale=0.4]{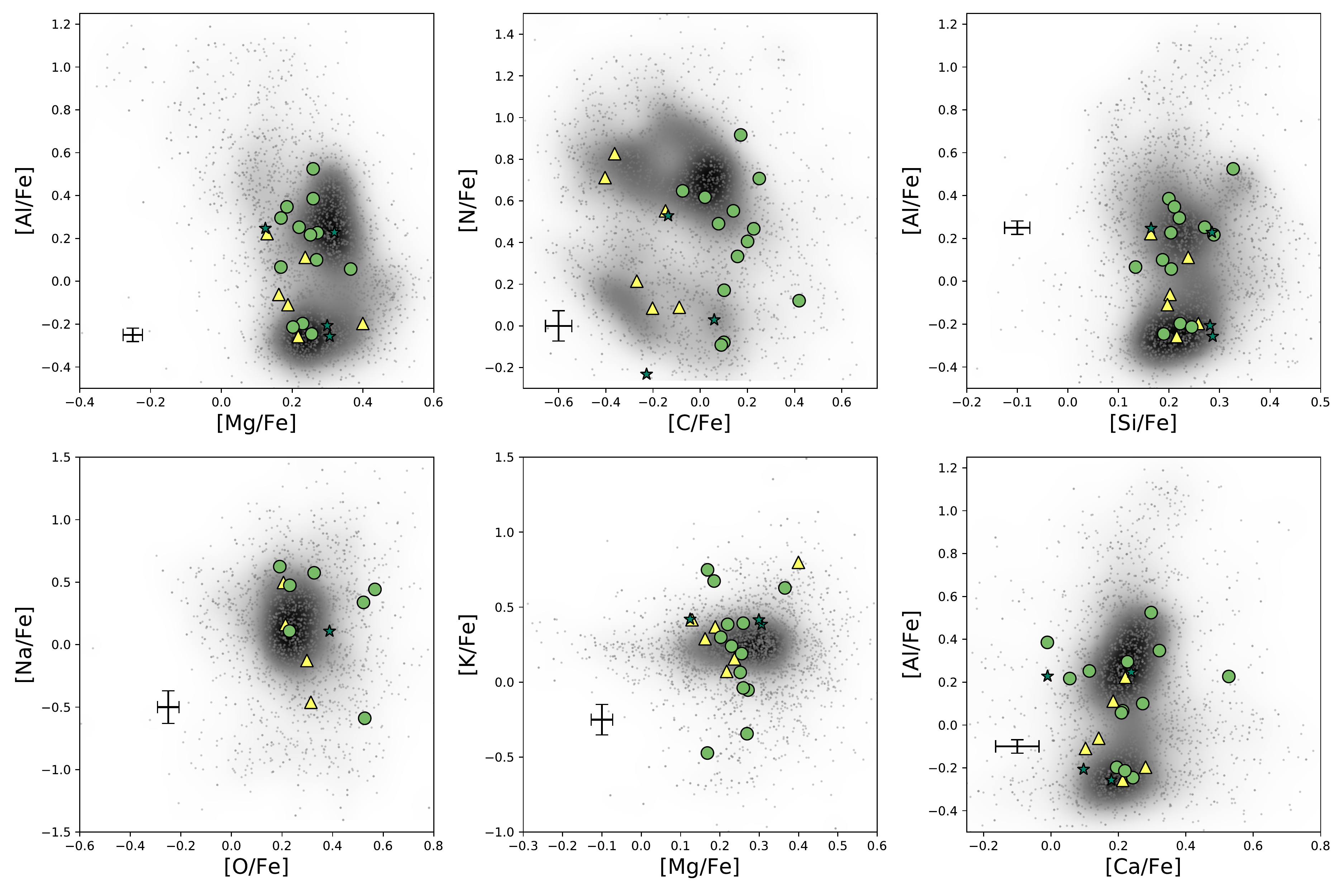}
    \caption{Known Mg-Al, C-N, and O-Na anticorrelations, along with other discussed abundance planes in section~\ref{sec:abundances}: Si-Al, Mg-K, and Ca-Al. 
    This Figure uses the same background and symbols described Figure~\ref{fig:all-vs-fe}, and for the background the matched APOGEE stars from the 
    \protect\cite{schiavon17}, \protect\cite{masseron19} and \protect\cite{meszaros20} analysed cluster stars. 
    We also apply a Gaussian smoothing to this sample, shown as the coloured background, to emphasise the high-density areas of the distributions.}
    \label{fig:abundances}
\end{figure*}

\subsubsection{The Mg-Al anticorrelation and the non-existent Si-leakage}

The Mg-Al chain is a complex set of proton capture nuclear reactions that need high temperatures (50 MK) to trigger 
\citep{prantzos07,carretta09b,cassisi12}. For this reason, the anticorrelation between Mg-Al that is naturally produced 
through this chain, cannot happen inside low mass stars. Massive stars are thought to be responsible for producing the 
observed anticorrelations, with some external mechanism needed to transport processed material to the surface of low mass 
RGB stars \citep{renzini15,bastian18}.

Upper left panel in Figure~\ref{fig:abundances} shows the comparison of our \mgfe\ and \alfe\ abundances with stars of 
other APOGEE observed GCs \citep{schiavon17,masseron19,meszaros20} from the Galactic bulge to the halo. For both catalogues, its 
stars were matched with the DR16, resulting in more than 680 stars in total. Overall the red-giant 
stars in NGC~6544 follow the same trend that other globular clusters at the same metallicity as M13, M2, and M3 with 
\feh\ equal to -1.53 dex, -1.65 dex, and -1.50 dex, respectively. For a cluster as low-mass as NGC~6544, an approximate 
mono abundance of \mgfe\ is expected. Despite that, we found that there is some spread around the median \mgfe\ value 
of $0.24 \pm 0.07$ dex. Nevertheless, this spread is well within two times the observational error, that also depends 
on the evolutionary stage marked with different symbols on the Figure. A clear sign of the Mg-Al anticorrelation is 
more evident in the \alfe\ abundances, in which we observe a high amplitude of almost $\Delta\alfe\sim 0.8$ dex. 
The same result was reported by \cite{carretta09b}, also observing a \mgfe-depletion for stars with \alfe\ $\geq 0.8$ dex 
in massive and/or metal-poor globular clusters. 
We report a maximum \alfe\ abundance of 0.52 dex, which agrees with the results of \cite{carretta09b}. 
There appears to be a gap around \alfe $= 0$ dex, and even AGB stars appear to split in two groups. Often, a large Al-spread 
and decrease in the \mgfe\ for stars with \alfe$\leq 0.8$ dex, could suggests a possible Si-leakage from the Mg-Al chain, that should be seen as 
an enhancement in the \sife\ abundance.
We do not detect any sign of \mgfe\ depletion. Indeed, although the Mg-Al knee is observable in the APOGEE data, none of the cluster members shows an evident Mg-poor, Al-rich behaviour.
In fact, there is only a few clusters observed with APOGEE in which this effect have been documented \citep[see][for more details]{masseron19,meszaros20}.
Another observable consequence of the Mg-depletion would be the increase of the \sife\ spread among individual stars in a GC, 
also known as Si-leakage, which is also not observed. 
This effect have been observed and described in depth in GCs more metal-poor than NGC~6544 \citep{meszaros20}, like NGC~6341/M92 (\feh$= -2.31$ dex) and NGC~7078/M15 (\feh$=-2.37$ dex).
Within the cluster, the \sife\ abundances remain constant, with a median value of $0.22 \pm 0.05$ dex, for cluster stars showing large differences in \alfe. 
Two groups of stars with different \alfe\ abundances can be identified, with separation around \alfe $\sim 0$ dex.
We will assign the labels of FG, and SG to the groups depending on the \alfe\ abundance lower, and greater than $0.0$ dex, respectively. 
These two groups, FG and SG, median \alfe\ values have a total difference of $0.42$ dex and a standard deviation of $0.23$ dex, without counting the AGB stars.
Both values agrees with the results of \cite{meszaros20} for other APOGEE observed clusters around the metallicity that we derive for NGC~6544.
Similarly, we can separate FG and SG stars in the C-N plane.

\subsubsection{The C-N anticorrelation}

The C-N anticorrelation is a natural consequence of the CNO cycle, which depletes C and O and enhances N in the 
H-burning \citep{cassisi12}. These products may be carried to the upper layers of the star either through some 
extra mixing phenomenon along the RGB, or because of pollution of the stellar atmosphere by the winds of high
mass AGB stars. The central upper panel in Figure~\ref{fig:abundances} shows the anticorrelation for NGC~6544 
in comparison with the globular clusters analysed by \cite{schiavon17}, \cite{masseron19}, and \cite{meszaros20}. 
Notice that we must be careful with the \cfe\ values derived by ASPCAP. The value represents the best fit value for the whole 
spectrum, and according to \cite{meszaros15}, only upper values can be measured below \feh\ $= -1.7$ dex. 
Fortunately, only two AGB stars fall under this metal-poor regime, and only one does not return any value in 
the DR16. A clear separation between pre and post-RGB-bump is seen through the \cfe abundances.

\subsubsection{The Na-O anticorrelation}

The Na-O anticorrelation is produced by oxygen and sodium, produced by the combined action of the CNO and Ne-Na 
cycles \citep{graton04, cassisi12}. While the CNO cycle decreases the O abundance (maintaining the C+N+O overall 
count), the Ne-Na cycle produces Na, creating the anticorrelation. The latter cycle is efficient in intermediate-mass 
stars during the thermal pulses of the AGB phase, but can also be active at the bottom of the convective envelope 
in RGB stars, since it requires lower temperatures compared with the Mg-Al cycle (around 40 MK).
Observationally, Na is difficult to measure in APOGEE spectra, because its two lines are very weak in the H-band.
That is the reason that all the \nafe\ quantities reported (and its associated errors) must be considered with caution, as the most uncertain values presented in this work.
On the same behaviour, O can be difficult to detect for metal-poor stars \citep[see the previous subsection, and][]{masseron19}.
As a result, only a few cluster stars have both elements measured, and are shown in the lower-left panel of 
Figure~\ref{fig:abundances}. As a broad overview, NGC~6544 stars follow the expected anticorrelation, like other 
APOGEE clusters, with a large \nafe\ spread of $\Delta\nafe\sim 1.1$ dex.
This is in agreement with the results from other clusters \citep{gratton04}, which also shows a larger spread in
\nafe\ with respect to \alfe.

\subsubsection{The Odd-Z element K}

Finally, we analyse an element that present strange patterns in the NGC~6544 APOGEE DR16 data.
Potassium, as an odd-Z element, is produced from Argon by a proton capture reaction. It needs a 
much higher temperature than other cycles reviewed in this section, with 120-180 MK to trigger 
\citep{ventura12, iliadis16}. There are available only a few lines in the APOGEE spectra, but as 
shown in \cite{masseron19}, \kfe was successfully recovered and reported a negligible K production 
over all the GCs analysed.

Contrary to the results for other GCs, we observe a broad \kfe\ spread, with up to $\sim 1.25$ dex in range.
This abundance variation matches the values reported by \citet{carretta13} and \cite{mucciarelli17} with high \kfe\ variance.
The central lower panel in Figure~\ref{fig:abundances} shows the Mg-K plane. As seen in the background 
for other GCs observed by APOGEE \citep{schiavon17, masseron19, meszaros20}, a constant value around solar 
composition was reported, i.e., no K-production was found.  
K-enhanced stars are indeed expected in globular clusters if the polluters achieve high temperatures to 
create Potassium. Like K-rich stars, subsequent generations will also show Mg-depleted and Si-enhanced 
stars due to the lower temperature barrier of the Mg-Al-Si cycle, as shown in \cite{meszaros20} for some 
APOGEE GC stars. 

With the present observations, we cannot determine the origin of the peculiar K-enhancement, 
which is not compatible with a constant value given the significant star-to-star variations. Optical 
follow-up is necessary to constraint the discrepancy extent of our measurements.

\subsubsection{The $\alpha$-element Ca}

Most of the Calcium that we can measure in GCs stars is produced by supernovae, and therefore not affected 
by the H-burning processes neither of the stars nor of the polluters. \cite{meszaros20} reported constant 
\cafe\ values for APOGEE GCs, though warning about the weakness of the Ca lines in the near-IR at low 
metallicities. Their results are shown in the lower right panel of Figure~\ref{fig:abundances} as background, 
along with the NGC~6544 stars. 

In general, cluster stars follow the overall trend of constant Calcium abundance, 
centred at \cafe$= 0.21 \pm 0.09$ dex. There is a deviation of this behaviour, but only for stars with 
\alfe$\lesssim 0.15$ dex. These four stars,  (3 lRGB and 1 AGB), however, also present large \cafe\ errors 
according to ASPCAP. If we do not take into account these four outliers, the Calcium abundance can be described 
as a constant, with stat-to-star deviations within the errors.

\subsubsection{ Neutron-capture elements: Ce and Nd}

Neutron capture elements, like Cerium and Neodymium, can be formed by a wide variety of channels in different contexts within the Galaxy \citep{sneden08}.
However in a GC, the possible scenarios that can occur can be limited to a few like a burning shell in a massive stars or an evolved 
AGB star in the thermally pulsating phase \citep{bisterzo11, bisterzo14}.
Within the APOGEE survey, multiple measurements of Ce and Nd have been reported since its characterisation and inclusion in subsequent DRs \citep{hasselquist16, cunha17}.
Given the limited range of temperatures in which Ce can be reliably fitted and measured \citep[\teff$<4400$ K according to][]{meszaros20}, 
all of the ASPCAP \cefe\ abundances reported here and in the DR16 not in this temperature range, can be considered only as upper limits. 
ASPCAP \cite{garciaperez16}, as an automatic abundance pipeline, cannot handle correctly the difference between an upper limit and a non-detection in which the line is absent from the spectra.
This is the case of all the Ce abundance measured and reported in this section.
Moreover, to better constraint the Ce abundances, the BACCHUS \citep{bacchus} code was used to properly account for the non-detections, 
even including more Ce lines to the fitting procedure\citep{masseron19,meszaros20}.

\begin{figure}
    \centering
    \includegraphics[scale=0.35]{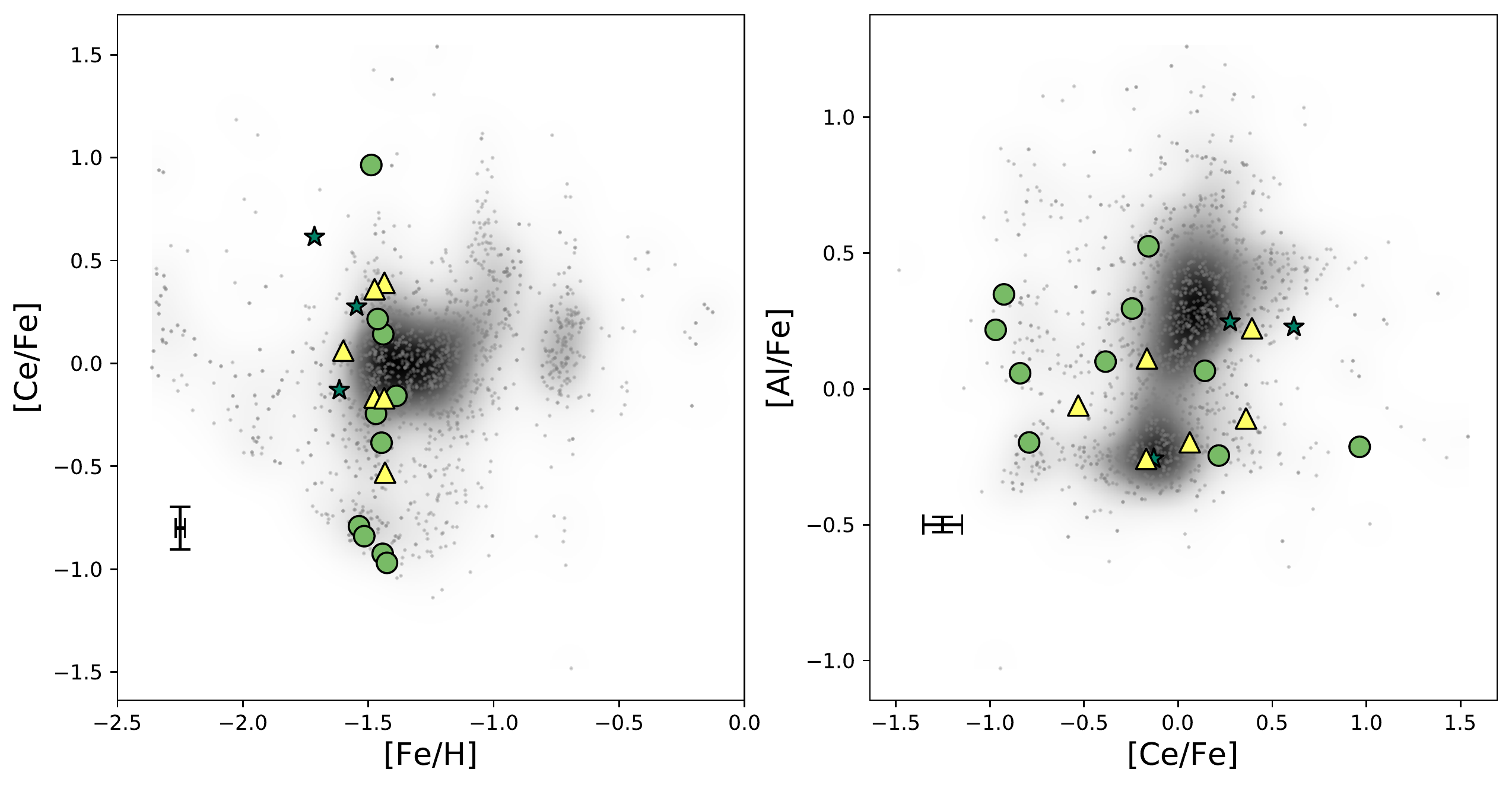}
    \caption{Upper limits for the \cefe\ detections vs \feh\ ({\bf Left panel}) and \alfe\ ({\bf Right panel}) for the APOGEE observed clusters present in the DR16. 
    NGC~6544 stars are colour-coded with the same markers as in Figure~\ref{fig:all-vs-fe} and \ref{fig:abundances}.
    A spread of almost $\sim2$ dex in \cefe\ can be explained by the inability of the ASPCAP pipeline to treat upper limits in the analysed spectra (see text). }
\label{fig:CeFe}    
\end{figure}

Given that information, we can explain the behaviour of the \cefe\ within the cluster.
Figure~\ref{fig:CeFe} shows the \cefe\ abundance as a function of the metallicity and \alfe\ for NGC~6544 and other APOGEE observed clusters. 
The wide range of \cefe\ of $1.9$ dex are solely explained by the fact that ASPCAP is not able to correctly fit the Ce line.
As showed in \cite{meszaros20}, $\omega$ Cen presents the most dramatic Ce enrichment of $\sim 1$ dex across $\sim 1$ dex of \feh\ enrichment, followed by NGC~1851/47Tuc.
This marked difference exemplifies how our measurements do not represent the actual state of NGC~6544 \cefe.
Further analysis with cooler stars within the cluster are necessary before reliably establish the Ce status of NGC~6544.

Finally, no \ndfe\ have been reported for the selected NGC~6544 stars due to the combination of \teff\ and \feh\ that the spectra of the GC presents.
According to \cite{hasselquist16}, only a fraction close to the $\sim20$\% of the total APOGEE spectra will show Nd variations that are strong enough to be detected at a high-SNR.

\subsection{Al-N correlation and its relation with NGC~6544 mass}

\begin{figure}
    \centering
    \includegraphics[scale=0.39]{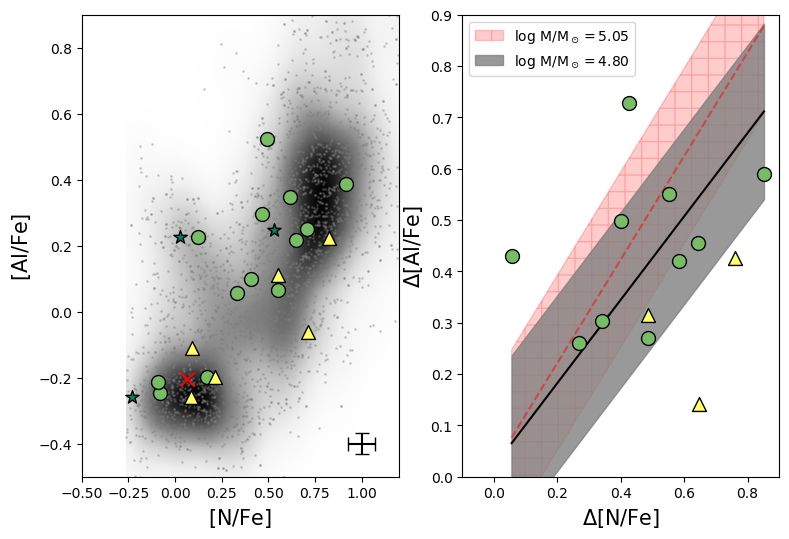}
    \caption{ ({\bf Left panel}): \alfe-\nfe\ correlation for the cluster stars. A clear separation between FG and SG at \nfe$< 0.4$ dex and \alfe$< 0$ dex for both abundances, respectively.
    Same symbols were used than in other figures for the different evolutionary stages. The red cross indicate the mean value of the FG stars, used later in the $\Delta$\xfe\ values.
    ({\bf Right panel}): $\Delta$\nfe-$\Delta$\alfe\ diagram as is shown in \protect\cite{nataf19}, with the continuous solid line representing the best fit using the lower mass value indicated in the legend,
    with its corresponding error as shaded area. Additionally, we include the fit as a dashed line (and hatched area for the error)  with the mass value used by \protect\cite{nataf19} to derive this relation. 
    Note that only SG stars are showed in this panel for which $\Delta$\xfe\ can be measured.}
\label{fig:AlFeNFe_nataf}    
\end{figure}

Finally, and with the aim to check the consistency of the previous APOGEE results, 
we perform the same analysis of \cite{nataf19} that relates the amplitude of the Al- and N-spread with the cluster mass. 
This relation was built in the past DR14 version of the APOGEE database, however considering that the calibrations within DRs consists only in shifts on the abundance plane, 
we can assume that is still valid within errors.
To built the relation, we need first to define the $\Delta$\xfe\ quantity, which is defined as

\begin{equation*}
\Delta\xfe_i = \xfe_{\rm Gen\ II,i} - \langle\xfe_{\rm Gen\ I}\rangle,
\end{equation*}
which relates the difference within the \xfe\ value of each $i$-star with SG abundance patterns with the mean value of \xfe\ but measured for FG stars.
As derived in equation~6 of \cite{nataf19}, we apply the relation within \feh\ and $\Delta$\nfe\ of the individual SG stars, and the cluster mass $\log{{\rm M/M_\odot}}$
to derive the total $\Delta$\alfe\ of the same stars. Given that the only unknown quantity is the total cluster mass, we can apply this relation to constraint the possible NGC~6544 mass values.
The procedure is performed by comparing the reason of both $\Delta$\alfe\ quantities, derived from Aluminum abundances, and from fixing the cluster mass. 
After calculating the median of all the stars, the most accurate mass will produce a value of $\Delta{\rm Al} = \Delta$\alfe$_{\rm spectra}/\Delta$\alfe$_{\rm fixed\ mass}$ closer to $1$.
The results shown in Figure~\ref{fig:AlFeNFe_nataf} correlates well with the statements of \cite{nataf19} in both the \alfe-\nfe\ correlation (left panel) and the $\Delta$\alfe-$\Delta$\nfe\ plane. 
Note that there are two different NGC~6544 total masses in the literature which differs by a factor of $\sim 2$, one listed in the \cite{harris96} catalogue 
and used by \cite{nataf19} of $\log{{\rm M/M_\odot}} = 5.05$, and the other one derived by \cite{baumgardt18} of $\log{{\rm M/M_\odot}} = 4.80$.
Also is important to highlight that at the time of fixing the cluster mass from the relation to calculate the trend line, 
we also fixed the metallicity of the cluster to \feh\ $= -1.44$ dex, as is stated in subsection~\ref{sec:fehalpha}. 
We use both masses and conclude that the \cite{baumgardt18} fully agrees within the errors the Al-N-Mass relation, which is showed as a continuous line in the right panel of Figure~\ref{fig:AlFeNFe_nataf}.
To quantify this statement, both fixed masses produces a value of $\Delta{\rm Al}$ equal to $0.82$ and $1.00$, for the \cite{nataf19} and \cite{baumgardt18} masses, respectively.

\section{Galactic context}
\label{sec:gcontext}

\begin{figure*}
    \centering
    \includegraphics[scale=0.5]{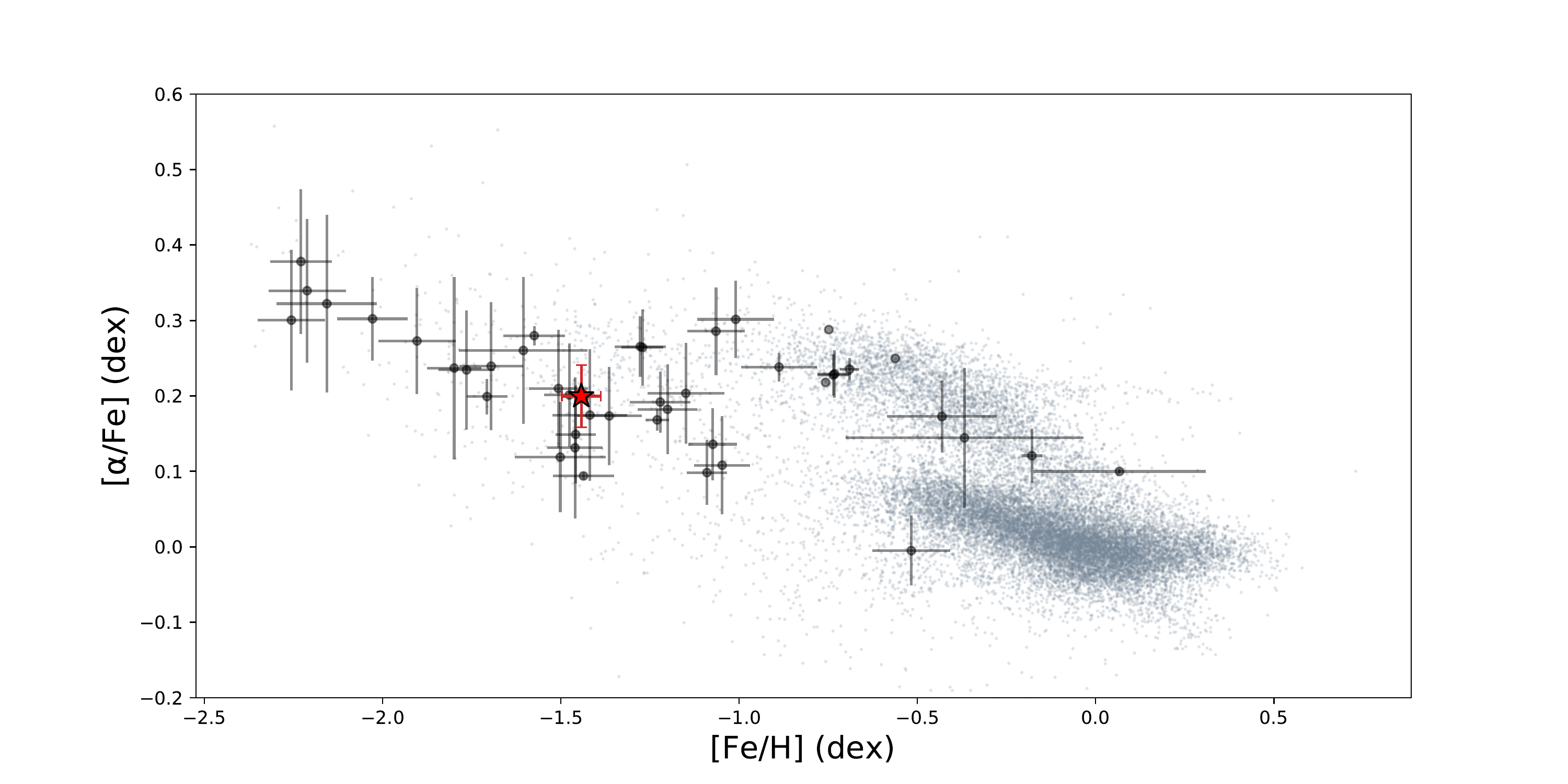}
    \caption{Tinsley diagram of the \alphafe enrichment across metallicity for the observed globular clusters by APOGEE \citep{schiavon17, masseron19, meszaros20}. 
    Despite being in a similar position of some halo globular clusters (NGC~5272/M~3, NGC~7089/M~2, and NGC~6205/M~13 with \feh\ = -1.49, -1.56, and -1.60 dex, respectively), 
    NGC~6544 (shown as a red star) was classified as thick disk GC by orbital reasons. As a background reference (small points), we place a 
    randomly selected 10\% of the whole APOGEE catalogue. The bulge, disk, and halo are visible, even showing the sequence of anomalous 
    stars at \alphafe\ $\sim$ 0.2 dex, described by \citet{zasowski19}.}
\label{fig:galcontext}    
\end{figure*}

NGC~6544 is projected on the sky towards the bulge area in a very crowded region at $\left( \ell, b\right) \sim \left(5.8^\circ, -2.2^\circ\right)$. 
The first comprehensive observations describe the cluster as a relatively metal-poor one,
unlike most other inner Galactic GCs \citep[][and clusters there in]{cohen14,cohen17,cohen18}.
Indeed, \cite{bica16} and \cite{contreras-ramos17} classify NGC~6544 as a halo intruder in the bulge region
based on metallicity estimates and simple orbital constraints. More recently, \cite{perezvillegas19} derived a 
more robust probability-based membership based on state of the art Galactic potentials for GCs in the inner Galaxy. 
With a very high percentage (within $\sim$80 to 97\% depending on the Galactic bar parameters), NGC~6544 was classified as thick disk GC.
Based on its \feh\ we can state that the cluster is consistent with the metal-poor tail of the canonical thick disk.
In order to contextualise the present results for the cluster, we built a catalogue of GCs observed by APOGEE 
\citep{schiavon17,masseron19, meszaros20}. The median metallicity and total $\alpha$-enhancement, as measured by ASPCAP in the DR16, were calculated. 
Figure~\ref{fig:galcontext} shows the chemical neighbourhood of NGC~6544, together with other clusters in the halo (NGC~7078/M~15, NGC~6341/M~92, 
NGC~5024/M~53, NGC~5466, NGC~7089/M~2, NGC~5272/M~3, NGC~6205/M~13 and NGC~5904/M~5), thick disk (NGC~6171/M~107, Pal 6, and NGC~6838/M~71), and bulge (NGC~6522, Ter 5, NGC~6528, and NGC~6553). 
We have also added a sample of randomly selected field stars from APOGEE DR16, representing 10\% of the total catalogue, 
as comparison. 

\begin{figure*}
    \centering
    \includegraphics[scale=0.4]{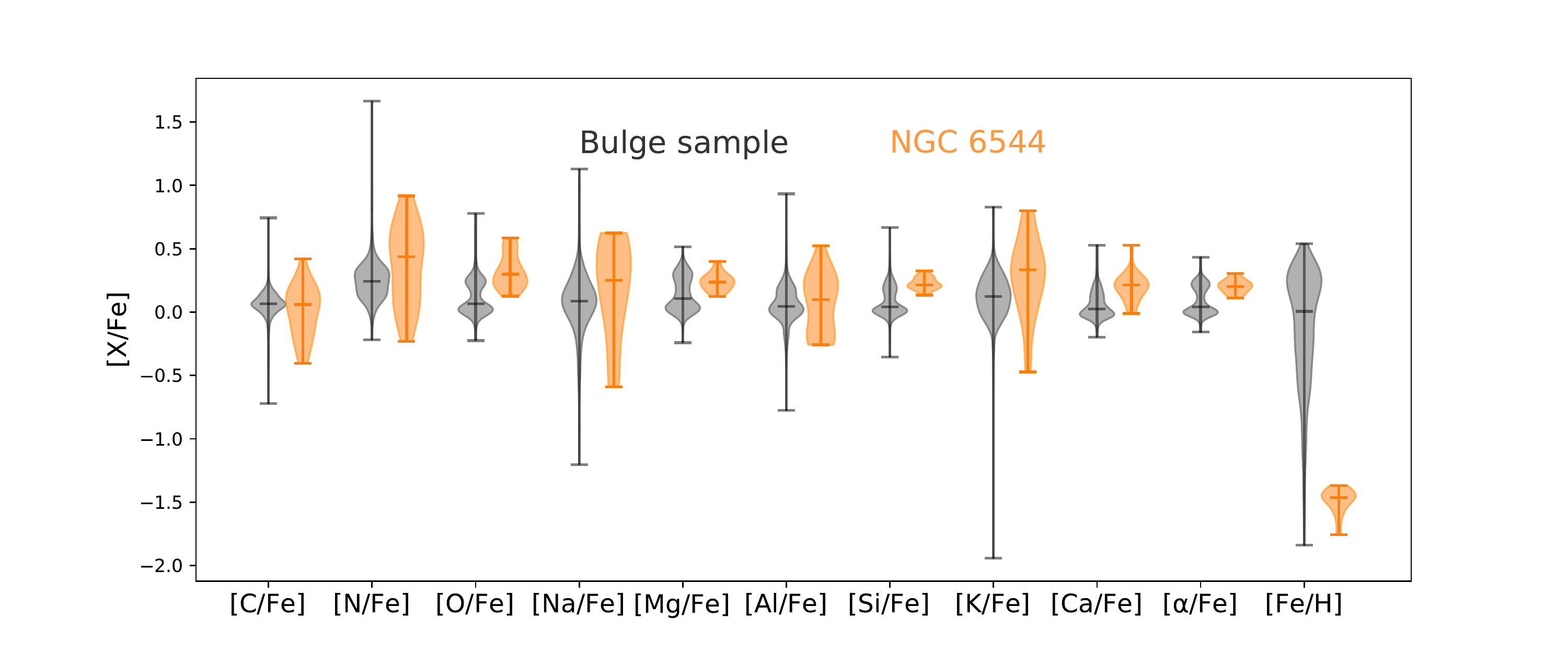}
    \caption{\xfe\ abundance density estimation comparison between bulge (left grey symbols) and 
    cluster (right orange symbols) stars. Each violin representation was normalised, and indicates with horizontal lines the median and limits of the distribution.}
    \label{fig:all-vs-median}
\end{figure*}

As a detailed comparison between the abundances of cluster stars and those of its spatial neighborhood, we show in 
Figure~\ref{fig:all-vs-median} the abundances of several elements in the cluster and in a sample of 3731 field stars
located within $2^\circ$ from the cluster centre, which are mostly bulge stars.
The normalised abundance distribution are shown as a density estimation for both bulge and cluster population,
whereby the median, min and max of the measurements of each element are marked. As it is clear from the Figure, 
cluster stars show a wider range of abundances than field stars, for all the elements except Mg and Si. The effect is especially
clear in \cfe, \nfe, \nafe, \alfe\ (with hints of the bimodality reported in section~\ref{sec:abundances}), and 
\kfe. Compared with previous results for field stars that may belong to GCs \citep{fernandeztrincado16, fernandeztrincado17},
we do not find any unusual abundance ratios in the nearby bulge stars that can be directly linked to NGC~6544. 

Finally, given the remarkable elongation displayed by NGC~6544 caused by the Galaxy \citep{contreras-ramos17}, 
we also search for extra tidal stars within the APOGEE DR16. We use the derived metallicity and space velocity 
for the cluster within this work as a first guess. Unfortunately, we did not find additional members within 
$\sim 10^\circ$ from the cluster centre. The use of more extended and deep data is mandatory to constrain the 
expected tidal tails of NGC~6544. Upcoming releases of the APOGEE data, and also forthcoming stellar wide-field 
multi-object spectroscopic surveys like 4MOST \citep{4MOST} and MOONS \citep{MOONS}, will help to identify extra 
tidal members, if they exist, as it was strongly suggested by previous studies \citep{cohen14}.

\section{Concluding remarks}
\label{sec:summary}

We have performed a near-IR high-resolution spectral analysis for the reddened GC NGC~6544. 
With 23 confirmed members in the APOGEE DR16 data, this is the most complete study up to date, to characterise the abundance patterns of this cluster.
Across the RGB (below and above the bump) and AGB, the abundances derived by the ASPCAP pipeline were presented
and discussed in the context of other Galactic GCs. Known anticorrelations (Mg-Al, C-N, Na-O) were used to isolate 
two generations of stars, a FG with 9 members and a SG with 14 members.
For a random targeting, these numbers are in very good agreement with predictions about the number of FG 
stars that still remain in GCs (close to $\sim 1/3$ of the total population). 

A large \alfe\ spread of $0.78$ dex, negligible \mgfe\ dispersion, a significant variation of CNO, across all the 
sampled evolutionary phases was a pivotal factor to divide different generations of stars.
The metallicity and $\alpha$-element abundances for RGB stars, from which we have a bonafide abundance determination, 
are $-1.44 \pm 0.05$ dex and $0.20 \pm 0.04$ dex, respectively. 
Using a correlation within the cluster Aluminium, Nitrogen, and metallicity we were able to constraint the NGC~6544 mass, 
favouring the less massive proposed scenario with $\log{{\rm M/M_\odot}} = 4.80$.

Multi-survey synergies were used to derive other cluster properties, such as the distance, reddening law towards the cluster and confirming its Galactic classification.
Finally, no extra tidal or chemically peculiar stars were found around the cluster, using data from the APOGEE DR16. 
Still, this picture could change with future and deeper spectroscopic surveys, which might explain the spatial elongation of NGC~6544. 

\section*{Acknowledgements}

Support for the authors is provided by the BASAL Center for Astrophysics 
and Associated Technologies (CATA)
through grant PFB-06, and the Ministry for the Economy, Development, and
Tourism, Programa Iniciativa Cientifica Milenio through grant IC120009,
awarded to the Millennium Institute of Astrophysics (MAS). 
This work is part of the Ph.D. thesis of F.G., funded by grant CONICYT-PCHA Doctorado Nacional 2017-21171485.

F.G. acknowledge support by CONICYT-PCHA Doctorado Nacional 2017-21171485.
ARA acknowledges support from FONDECYT through grant 3180203.
We acknowledge support from FONDECYT Regular grants No. 1150345 (MZ and FG), and No. 1170121 (D. M.).
DAGH acknowledges support from the State Research Agency (AEI) of the Spanish
Ministry of Science, Innovation and Universities (MCIU) and the European
Regional Development Fund (FEDER) under grant AYA2017-88254-P.
SV gratefully acknowledges the support provided by Fondecyt reg. n. 1170518. 
SV gratefully acknowledges the support from the Chilean
BASAL Centro de Excelencia en Astrof\'isica y Tecnolog\'ias Afines (CATA)
grant AFB-170002. J.G.F-T is supported by FONDECYT No. 3180210 and 
Becas Iberoam\'erica Investigador 2019, Banco Santander Chile.
D.G. gratefully acknowledges support from the Chilean Centro de Excelencia en Astrof\'isica
y Tecnolog\'ias Afines (CATA) BASAL grant AFB-170002.
D.G. also acknowledges financial support from the Direcci\'on de Investigaci\'on y Desarrollo de
la Universidad de La Serena through the Programa de Incentivo a la Investigaci\'n de
Acad\'emicos (PIA-DIDULS).\\

We gratefully acknowledge the use of data from the VVV/VVVx ESO Public Survey
program ID 179.B-2002/198.B-2004 taken with the VISTA telescope, and data products
from the Cambridge Astronomical Survey Unit (CASU). The VVV Survey data
are made public at the ESO Archive.\\

Funding for the Sloan Digital Sky Survey IV has been provided by
the Alfred P. Sloan Foundation, the US Department of Energy Office of
Science, and the Participating Institutions. SDSS-IV acknowledges
support and resources from the Center for High-Performance Computing at
the University of Utah. The SDSS website is www.sdss.org.\\

SDSS-IV is managed by the Astrophysical Research Consortium for the
Participating Institutions of the SDSS Collaboration including the
Brazilian Participation Group, the Carnegie Institution for Science,
Carnegie Mellon University, the Chilean Participation Group, the
French Participation Group, Harvard-Smithsonian Center for Astrophysics,
Instituto de Astrof\'isica de Canarias, The Johns Hopkins University,
Kavli Institute for the Physics and Mathematics of the Universe (IPMU)
University of Tokyo, Lawrence Berkeley National Laboratory, Leibniz
Institut f\"ur Astrophysik Potsdam (AIP),  Max-Planck-Institut
f\"ur Astronomie (MPIA Heidelberg), Max-Planck-Institut f\"ur
Astrophysik (MPA Garching), Max-Planck-Institut f\"ur
Extraterrestrische Physik (MPE), National Astronomical
Observatory of China, New Mexico State University,
New York University, University of Notre Dame,
Observat\'ario Nacional/MCTI, The Ohio State University,
Pennsylvania State University, Shanghai Astronomical Observatory,
United Kingdom Participation Group, Universidad Nacional
Aut\'onoma de M\'exico, University of Arizona, University of Colorado
Boulder, University of Oxford, University of Portsmouth, University of
Utah, University of Virginia, University of Washington, University of
Wisconsin, Vanderbilt University, and Yale University.\\

This work has made use of data from the European Space Agency (ESA) mission
{\it Gaia} (\url{https://www.cosmos.esa.int/gaia}), processed by the {\it Gaia}
Data Processing and Analysis Consortium (DPAC,\url{https://www.cosmos.esa.int/web/gaia/dpac/consortium}). 
Funding for the DPAC has been provided by national institutions, in particular the institutions
participating in the {\it Gaia} Multilateral Agreement. \\

This publication makes use of data products from the Wide-field Infrared Survey Explorer, 
which is a joint project of the University of California, Los Angeles, and the Jet Propulsion Laboratory/California Institute of Technology, 
and NEOWISE, which is a project of the Jet Propulsion Laboratory/California Institute of Technology. 
WISE and NEOWISE are funded by the National Aeronautics and Space Administration. \\

This paper uses data from the MW-BULGE-PSFPHOT compilation \cite{surot19,surot20}.
This research has made use of the services of the ESO Science Archive Facility.

This publication makes use of data products from the Two Micron All Sky Survey, which is a joint 
project of the University of Massachusetts and the Infrared Processing and Analysis Center/California 
Institute of Technology, funded by the National Aeronautics and Space Administration and the National Science Foundation.\\

This research made use of: TOPCAT \citep{topcat}; pandas \cite{pandas}; IPython \citep{ipython}; numpy \citep{numpy}; matplotlib \citep{matplotlib};
Astropy, a community-developed core Python package for Astronomy \citep{astropy, astropy2}; dust\_extinction, an Astropy-affiliate package; and scikit-learn \cite{scikit-learn}.
This research has made use of NASA's Astrophysics Data System.

%%%%%%%%%%%%%%%%%%%%%%%%%%%%%%%%%%%%%%%%%%%%%%%%%%

\section*{Data Availability}

All APOGEE DR16 data used in this study is publicly available and can be found at: \href{https://www.sdss.org/dr16/}{https://www.sdss.org/dr16/}.
However we prepared Table~\ref{tab:all} as a supplemental material in the journal webpage.



\bibliographystyle{mnras}
\bibliography{NGC6544} % if your bibtex file is called example.bib

%%%%%%%%%%%%%%%%%%%%%%%%%%%%%%%%%%%%%%%%%%%%%%%%%%

%%%%%%%%%%%%%%%%% APPENDICES %%%%%%%%%%%%%%%%%%%%%

\appendix

\section{Appendix}

\begin{landscape}
\begin{table}
  \centering
  \caption{Identifiers and stellar parameters, \alphafe, \feh\ and elemental abundances from the NGC~6544 members. Median values, standard deviation and mean ASPCAP errors are also derived.}
  \label{tab:all}
\scriptsize
\begin{tabular}{c|c|c|c|c|c|c|c|c|c|c|c|c|c|c|c|c|c}
APOGEE ID & Type & \teff & \logg  & \vhelio  &  SNR  & \feh  & \alphafe & \cfe  & \nfe  & \ofe  & \nafe & \mgfe & \alfe & \sife & \kfe & \cafe & \cefe  \\
          &      &  (K)  &  (dex) &  (\kms)  &       & (dex) &   (dex)  & (dex) & (dex) & (dex) & (dex) & (dex) & (dex) & (dex) & (dex) & (dex) & (dex) \\
\hline
2M18070312-2501429 & lrgb & 4949.61 & 2.42 & -41.9  & 119.35 & -1.47 & 0.16 &  0.23 &  0.47 & 0.23 &  0.11 & 0.17 &  0.30 & 0.22 &  0.75 &  0.23 & -0.25 \\
2M18070657-2500417 & agb  & 4797.31 & 1.49 & -30.9  & 291.71 & -1.55 & 0.11 & -0.14 &  0.53 & 0.21 &  ---  & 0.12 &  0.25 & 0.16 &  0.42 &  0.24 &  0.28 \\
2M18071190-2458586 & lrgb & 4929.79 & 2.27 & -39.2  &  70.09 & -1.41 & 0.23 &  0.42 &  0.12 & 0.52 &  0.34 & 0.27 &  0.23 & 0.20 & -0.05 &  0.53 &  ---  \\
2M18071347-2458525 & urgb & 4782.61 & 1.95 & -39.7  & 169.30 & -1.48 & 0.18 & -0.15 &  0.55 & 0.21 &  0.50 & 0.24 &  0.11 & 0.24 &  0.16 &  0.18 & -0.17 \\
2M18071474-2454595 & lrgb & 4943.43 & 2.17 & -37.9  & 205.98 & -1.54 & 0.26 &  0.10 &  0.17 & 0.57 &  0.44 & 0.23 & -0.20 & 0.22 &  0.24 &  0.19 & -0.79 \\
2M18071492-2457397 & agb  & 4819.18 & 1.71 & -34.2  & 152.02 & -1.62 & 0.29 & -0.23 & -0.23 & 0.28 &  ---  & 0.31 & -0.26 & 0.29 &  0.39 &  0.18 & -0.13 \\
2M18071622-2457231 & urgb & 4688.18 & 1.76 & -51.4  & 206.36 & -1.60 & 0.27 & -0.27 &  0.21 & 0.37 &  ---  & 0.40 & -0.20 & 0.26 &  0.80 &  0.28 &  0.06 \\
2M18071727-2504387 & lrgb & 5018.44 & 2.42 & -36.6  &  94.82 & -1.44 & 0.14 &  0.14 &  0.55 & 0.33 &  0.57 & 0.17 &  0.07 & 0.13 & -0.47 &  0.21 &  0.14 \\
2M18071947-2500517 & lrgb & 4986.75 & 2.27 & -50.6  & 224.29 & -1.46 & 0.23 &  0.10 & -0.08 & 0.53 & -0.59 & 0.26 & -0.25 & 0.19 &  0.19 &  0.24 &  0.22 \\
2M18072003-2502407 & lrgb & 5074.18 & 2.50 & -38.5  & 114.17 & -1.52 & 0.24 &  0.16 &  0.33 & 0.50 &  ---  & 0.37 &  0.06 & 0.20 &  0.63 &  0.21 & -0.84 \\
2M18072114-2458359 & lrgb & 4861.69 & 2.23 & -37.7  & 115.97 & -1.39 & 0.20 &  0.08 &  0.49 & 0.19 &  0.62 & 0.26 &  0.52 & 0.33 &  0.39 &  0.30 & -0.16 \\
2M18072123-2502436 & lrgb & 5112.84 & 2.65 & -30.5  & 117.47 & -1.37 & 0.16 &  0.17 &  0.92 & ---  &  ---  & 0.26 &  0.39 & 0.20 & -0.04 & -0.01 &  ---  \\
2M18072204-2501100 & urgb & 4674.79 & 1.73 & -37.5  & 159.89 & -1.44 & 0.21 & -0.20 &  0.09 & 0.30 & -0.13 & 0.22 & -0.26 & 0.22 &  0.07 &  0.21 & -0.17 \\
2M18072226-2500325 & urgb & 4470.25 & 1.39 & -38.2  & 492.25 & -1.44 & 0.13 & -0.36 &  0.83 & 0.19 &  ---  & 0.13 &  0.22 & 0.16 &  0.42 &  0.22 &  0.39 \\
2M18072230-2458509 & agb  & 4896.84 & 1.74 & -33.4  & 323.80 & -1.76 & 0.31 & -0.03 &  ---  & 0.39 &  0.11 & 0.30 & -0.21 & 0.28 &  0.42 &  0.10 &  ---  \\
2M18072402-2458088 & lrgb & 5099.30 & 2.51 & -37.2  &  79.06 & -1.37 & 0.20 &  0.25 &  0.71 & ---  &  ---  & 0.22 &  0.25 & 0.27 &  0.39 &  0.11 &  ---  \\
2M18072492-2458004 & agb  & 5062.54 & 1.90 & -36.8  &  82.36 & -1.71 & 0.21 &  0.06 &  0.03 & 0.21 &  ---  & 0.32 &  0.23 & 0.29 &  ---  & -0.01 &  0.61 \\
2M18072612-2457497 & lrgb & 4972.73 & 2.32 & -42.4  &  99.79 & -1.45 & 0.18 &  0.20 &  0.41 & 0.33 &  ---  & 0.27 &  0.10 & 0.19 & -0.34 &  0.27 & -0.39 \\
2M18072927-2502396 & lrgb & 4912.17 & 2.20 & -40.2  & 109.95 & -1.44 & 0.13 &  0.02 &  0.62 & 0.13 &  ---  & 0.19 &  0.35 & 0.21 &  0.67 &  0.32 & -0.93 \\
2M18072970-2456359 & urgb & 4316.64 & 1.19 & -47.1  & 650.68 & -1.43 & 0.19 & -0.40 &  0.71 & 0.21 &  0.16 & 0.16 & -0.06 & 0.20 &  0.29 &  0.14 & -0.53 \\
2M18073271-2500281 & lrgb & 4885.97 & 2.16 & -40.1  &  72.44 & -1.43 & 0.23 & -0.07 &  0.65 & 0.23 &  0.47 & 0.25 &  0.22 & 0.29 &  0.07 &  0.06 & -0.97 \\
2M18073312-2457396 & lrgb & 5029.56 & 2.35 & -40.7  & 129.57 & -1.49 & 0.24 &  0.09 & -0.09 & 0.58 &  ---  & 0.20 & -0.21 & 0.24 &  0.30 &  0.22 &  0.96 \\
2M18074298-2452101 & urgb & 4756.59 & 1.85 & -36.7  & 226.44 & -1.47 & 0.20 & -0.09 &  0.09 & 0.31 & -0.46 & 0.19 & -0.11 & 0.20 &  0.37 &  0.10 &  0.36 \\
\hline
Cluster median     &      &  ---    & ---  & -38.17 & 129.57 & -1.46 & 0.20 &  0.06 &  0.44 & 0.30 &  0.25 & 0.24 &  0.10 & 0.22 &  0.33 &  0.21 & -0.16 \\
Cluster 1$\sigma$  &      &  ---    & ---  &   5.10 & 138.04 &  0.10 & 0.05 &  0.20 &  0.31 & 0.14 &  0.38 & 0.07 &  0.23 & 0.05 &  0.31 &  0.11 &  0.52 \\
ASPCAP mean error  &      & 113.57  & 0.09 &   0.06 &  ---   &  0.02 & 0.02 &  0.05 &  0.07 & 0.05 &  0.13 & 0.03 &  0.03 & 0.02 &  0.10 &  0.06 &  0.11 \\
\hline
\end{tabular}
\end{table}
\end{landscape}

% Don't change these lines
\bsp    % typesetting comment
\label{lastpage}
\end{document}